\def\ThisFile{MarkovStructureOfGenomes.tex}
\def\ThisFileDate{2021/12/24}
\documentclass[11pt]{article}
\usepackage{ifthen,pifont,latexsym}
\usepackage[T1]{fontenc}
\usepackage{color}
\usepackage{times,avant}

\usepackage[round,authoryear]{natbib}
\usepackage{amsmath}
\usepackage{setspace}
\usepackage[pdftex]{graphicx}
{\end{description}}
{\end{itemize}}
{\end{enumerate}}
\def\Includefigs{true} 
\def\Thisfile{acronyms.tex}
\def\Thisfiledate{2021/08/13}
\def\NISS{National Institute of Statistical Sciences (NISS)
    \gdef\NISS{NISS}}

\def\NCSU{North Carolina State University (NCSU)
    \gdef\NCSU{NCSU}}
\def\UNC{University of North Carolina at Chapel Hill (UNC)
    \gdef\UNC{UNC}}
\def\CMU{Carnegie Mellon University (CMU)
    \gdef\CMU{CMU}}
\def\DU{Duke University (Duke)
    \gdef\DU{Duke}}
\def\UMI{University of Michigan (UMI)
    \gdef\UMI{UMi}}
\def\UMD{University of Maryland College Park (UMD)
    \gdef\UMD{UMd}}
\def\PU{Purdue University (Purdue)
    \gdef\PU{Purdue}}
\def\SMU{Southern Methodist University (SMU)
    \gdef\SMU{SMU}}
\def\GMU{George Mason University (GMU)
    \gdef\GMU{GMU}}
\def\UIC{University of Illinois at Chicago (UIC)
    \gdef\UIC{UIC}}
\def\LANL{Los Alamos National Laboratory (LANL)
    \gdef\LANL{LANL}}
\def\PNNL{Pacific Northwest National Laboratory (PNNL)
    \gdef\PNNL{PNNL}}
\def\GM{General Motors (GM)
    \gdef\GM{GM}}
\def\GSK{GlaxoSmithKline (GSK)
    \gdef\GSK{GSK}}
\def\VI{Visual Insights (VI)
    \gdef\VI{VI}}
\def\EIA{Energy Information Administration (EIA)
    \gdef\EIA{EIA}}
\def\EPA{Environmental Protection Agency (EPA)
    \gdef\EPA{EPA}}
\def\NCES{National Center for Education Statistics (NCES)
    \gdef\NCES{NCES}}
\def\BTS{Bureau of Transportation Statistics (BTS)
    \gdef\BTS{BTS}}
\def\BLS{Bureau of Labor Statistics (BLS)
    \gdef\BLS{BLS}}
\def\NCHS{National Center for Health Statistics (NCHS)
    \gdef\NCHS{NCHS}}
\def\BC{Census Bureau (Census)
    \gdef\BC{Census}}
\def\CB{Census Bureau (Census)
    \gdef\CB{Census}}
\def\NASS{National Agricultural Statistics Service (NASS)
    \gdef\NASS{NASS}}
\def\NSF{National Science Foundation (NSF)
    \gdef\NSF{NSF}}
\def\DMS{Division of Mathematical Sciences (DMS)
    \gdef\DMS{DMS}}
\def\CISE{Computer and Information Sciences and Engineering 
    (CISE)\gdef\CISE{CISE}}
\def\CATS{Committee on Theoretical and Applied Statistics (CATS)
    \gdef\CATS{CATS}}
\def\NRC{National Research Council (NRC)
    \gdef\NRC{NRC}}
\def\CNSTAT{Committee on National Statistics (CNSTAT)
    \gdef\CNSTAT{CNSTAT}}
\def\DOD{US Department of Defense (DoD)
    \gdef\DOD{DoD}}
\def\USGS{US Geological Survey (USGS)
    \gdef\USGS{USGS}}
\def\OMB{Office of Management and Budget (OMB)
    \gdef\OMB{OMB}}
\def\NSA{National Security Agency (NSA)
    \gdef\NSA{NSA}}
\def\DHS{Department of Homeland Security (DHS)
    \gdef\DHS{DHS}}
\def\CDC{Centers for Disease Control and Prevention (CDC)
    \gdef\CDC{CDC}}
\def\DARPA{Defense Advanced Research Projects Agency (DARPA)
    \gdef\DARPA{DARPA}}
\def\DOE{Department of Energy (DOE)
    \gdef\DOE{DOE}}
\def\CIA{Central Intelligence Agency (CIA)
    \gdef\CIA{CIA}}
\def\DTRA{Defense Threat Reduction Agency (DTRA)
    \gdef\DTRA{DTRA}}
\def\NIST{National Institute of Standards and Technology (NIST)
    \gdef\NIST{NIST}}
\def\NIAAA{National Institute on Alcohol Abuse and Alcoholism
    (NIAAA)
    \gdef\NIAAA{NIAAA}}
\def\ARO{Army Research Office (ARO)
    \gdef\ARO{ARO}}
\def\FDA{Food and Drug Administration (FDA)
    \gdef\FDA{FDA}}
\def\SAMSI{Statistical and Applied Mathematical Sciences
    Institute (SAMSI)\gdef\SAMSI{SAMSI}}

\def\NCDOT{North Carolina Department of Transporation (NCDOT)
    \gdef\NCDOT{NCDOT}}
\def\NCGBC{North Carolina Bioinformatics and Genomics Consortium 
    (NCGBC)\gdef\NCGBC{NCGBC}}
\def\RTI{RTI International (RTI)
    \gdef\RTI{RTI}}
\def\CIIT{CIIT Centers for Health Research (CIIT)
    \gdef\CIIT{CIIT}}
\def\DQRI{Data Quality Research Institute (DQRI)
    \gdef\DQRI{DQRI}}
\def\DIMACS{Center for Discrete Mathematics and Theoretical %
    Computer Science (DIMACS)\gdef\DIMACS{DIMACS}}
\def\HDF{Hereditary Disease Foundation (HDF)
    \gdef\HDF{HDF}}
\def\NCDM{National Center for Data Mining (NCDM)
    \gdef\NCDM{NCDM}}
\def\RTP{Research Triangle Park (RTP)
    \gdef\RTP{RTP}}
\def\ITDB{Intermodal Transportation Database (ITDB)
    \gdef\ITDB{ITDB}}
\def\TRI{Toxic Release Inventory (TRI)
    \gdef\TRI{TRI}}
\def\CPS{Current Population Survey (CPS)
    \gdef\CPS{CPS}}
\def\SASS{Schools and Staffing Survey (SASS)
    \gdef\SASS{SASS}}
\def\ITR{Information Technology Research (ITR)
    \gdef\ITR{ITR}}
\def\DC{data confidentiality (DC)
    \gdef\DC{DC}}
\def\DQ{data quality (DQ)
    \gdef\DQ{DQ}}
\def\DI{data integration (DI)
    \gdef\DI{DI}}
\def\IT{information technology (IT)
    \gdef\IT{IT}}
\def\SDL{statistical disclosure limitation (SDL)
    \gdef\SDL{SDL}}
\def\IQ{information quality (IQ)
    \gdef\IQ{IQ}}
\def\MCMC{Markov chain Monte Carlo (MCMC)
    \gdef\MCMC{MCMC}}
\def\CSV{comma-separated value (CSV)
    \gdef\CSV{CSV}}

\def\RMI{remote method invocation (RMI)
    \gdef\RMI{RMI}}
\def\SOAP{simple object access protocol (SOAP)
    \gdef\SOAP{SOAP}}
\def\XML{extensible markup language (XML)
    \gdef\XML{XML}}

\def\NHTSA{National Highway Traffic Safety Administration (NHTSA)%
    \gdef\NHTSA{NHTSA}}
\def\RDB{relational database (RDB)
    \gdef\RDB{RDB}\gdef\RDBMS{RBDMS}\gdef\RDBMSs{RDBMSs}}
\def\RDBMS{relational database management system (RDBMS)
    \gdef\RDB{RDB}\gdef\RDBMS{RBDMS}\gdef\RDBMSs{RDBMSs}}
\def\RDBMSs{relational database management systems (RDBMSs)
    \gdef\RDB{RDB}\gdef\RDBMS{RBDMS}\gdef\RDBMSs{RDBMSs}}
\def\GIS{geographical information system (GIS)\gdef\GIS{GIS}}
\def\DQTK{data quality toolkit (DQTK)
    \gdef\DQTK{DQTK}}
\def\DQRC{data quality report card (DQRC)
    \gdef\DQRC{DQRC}}
\def\TDQM{Total Data Quality Management (TQDM)
    \gdef\TDQM{TDQM}}
\def\OTR{optimal tabular release (OTR)
    \gdef\OTR{OTR}\gdef\OTRs{OTRs}}
\def\OTRs{optimal tabular releases (OTRs)
    \gdef\OTR{OTR}\gdef\OTRs{OTRs}}
\def\GUI{graphical user interface (GUI)
    \gdef\GUI{GUI}}
\def\OLTP{On-Line Transaction Processing (OLTP)
    \gdef\OLTP{OLTP}}
\def\GPRA{Government Performance Results Act (GPRA)
    \gdef\GPRA{GPRA}}

\def\CRM{Customer Relationship Management (CRM)
    \gdef\CRM{CRM}}
\def\HCI{human--computer interaction (HCI)
    \gdef\HCI{HCI}}
\def\NDHS{National Defense and Homeland Security (NDHS)
    \gdef\NDHS{NDHS}}
\def\MMR04{2004 Conference on Mathematical Methods in Reliability (MMR 2004)
    \gdef\MMR04{MMR 2004}}

\def\NCLB{No Child Left Behind Act (NCLB)
    \gdef\NCLB{NCLB}}
\def\CCD{Common Core of Data (CCD)
    \gdef\CCD{CCD}}

\def\NIH{National Institutes of Health (NIH)
    \gdef\NIH{NIH}}
\def\EFF{Electronic Frontier Foundation (EFF)
    \gdef\EFF{EFF}}
\def\GCD{graduation, completion and dropout (GCD)
    \gdef\GCD{GCD}}

\def\NCAR{National Center for Atmospheric Research (NCAR)
    \gdef\NCAR{NCAR}}

\def\TIMSS{Third International Mathematics and Science Study (TIMSS)
    \gdef\TIMSS{TIMSS}}
\def\PISA{Program for International Student Assessment (PISA)
    \gdef\PISA{PISA}}
\def\IEA{International Association for the Evaluation of
        Educational Achievement (IEA)
    \gdef\IEA{IEA}}
\def\NAEP{National Assessment of Educational Progress (NAEP)
    \gdef\NAEP{NAEP}}
\def\PIRLS{Progress in International Reading Literacy Study (PIRLS)
    \gdef\PIRLS{PIRLS}}

\def\OECD{Organization for Economic Cooperation and Development (OECD)
    \gdef\OECD{OECD}}

\def\GDC{graduation, dropout and completion (GDC)
    \gdef\GDC{GDC}}

\def\SMPC{secure multi-party computation (SMPC)
    \gdef\SMPC{SMPC}}
\def\HIPAA{Health Insurance Privacy and Accountability Act (HIPAA)
    \gdef\HIPAA{HIPAA}}

\def\ACS{American Community Survey (ACS)
    \gdef\ACS{ACS}}

\def\ESSI{Education Statistics Services Institute (ESSI)
    \gdef\ESSI{ESSI}}

\def\SCDM{Society for Clinical Data Management (SCDM)
    \gdef\SCDM{SCDM}}
\def\ACDM{Association for Clinical Data Management (ACDM)
    \gdef\ACDM{ACDM}}

\def\ASA{American Statistical Association (ASA)
    \gdef\ASA{ASA}}

\def\NCAA{National Collegiate Athletic Association (NCAA)
    \gdef\NCAA{NCAA}}
\def\GED{General Education Development (GED)
    \gdef\GED{GED}}

\def\ISU{Iowa State University (ISU)
    \gdef\ISU{ISU}}

\def\FIPS{Federal Information Processing System (FIPS)
    \gdef\FIPS{FIPS}}
\def\GSA{General Services Administration (GSA)
    \gdef\GSA{GSA}}

\def\AIR{American Institutes for Research (AIR)
    \gdef\AIR{AIR}}
\def\ESSIS{Education Statistics Services Institute---Statistics
    (ESSI--Stat)\gdef\ESSIS{ESSI-Stat}}

\def\NESSI{NAEP Education Statistics Services Institute (NESSI)
    \gdef\NESSI{NESSI}}

\def\ACM{Association for Computing Machinery (ACM)
    \gdef\ACM{ACM}}
\def\IEEE{Institute of Electrical and Electronics Engineers (IEEE)
    \gdef\IEEE{IEEE}}
\gdef\SIAM{Society for Industrial and Applied Mathematics (SIAM)
    \gdef\SIAM{SIAM}}
\def\IAOS{ISI Section on Official Statistics (IAOS)
    \gdef\IAOS{IAOS}}
\def\ISBA{International Society for Bayesian Analysis (ISBA)
    \gdef\ISBA{ISBA}}

\def\CDAC{Confidentiality and Data Access Committee (CDAC)
    \gdef\CDAC{CDAC}}
\def\CSIRO{Commonwealth Scientific and Industrial Research
    Organisation (CSIRO)\gdef\CSIRO{CSIRO}}
\def\TUCASI{Triangle Universities Center for Advanced Studies,
    Inc.\ (TUCASI)\gdef\TUCASI{TUCASI}}
\def\NCI{National Cancer Institute (NCI)
    \gdef\NCI{NCI}}

\def\BMSA{Board on Mathematical Sciences and their Applications
    (BMSA)\gdef\BMSA{BMSA}}

\def\NSCAW{National Survey of Child and Adolescent Well-Being
    (NSCAW)\def\NSCAW{NSCAW}}

\def\DAS{Data Analysis System (DAS)
    \gdef\DAS{DAS}\gdef\DASs{DAS's}}
\def\DASs{Data Analysis Systems (DAS's)
    \gdef\DAS{DAS}\gdef\DAS{DAS's}}

\def\CFFR{Committee on Federally Funded Research (CFFR)
    \gdef\CFFR{CFFR}}

\def\AJS{American Judicature Society (AJS)
    \gdef\AJS{AJS}}
\def\ECCR{Exploratory Center for Cheminformatics Research (ECCR)
    \gdef\ECCR{ECCR}}
\def\AAAS{American Association for the Advancement of Science (AAAS)
    \gdef\AAAS{AAAS}}

\def\RTF{Research Triangle Foundation (RTF)
    \gdef\RTF{RTF}}

\def\CFFR{Committee on Federally Funded Research (CFFR)%
    \gdef\CFFR{CFFR}}
\def\MPS{Directorate for Mathematical and Physical Sciences (MPS)%
    \gdef\MPS{MPS}}
\def\CMG{Collaborations in Mathematical Geosciences (CMG)%
    \gdef\CMG{CMG}}

\def\IMS{Institute of Mathematical Statistics (IMS)
    \gdef\IMS{IMS}}
\def\ISI{International Statistical Institute (ISI)
    \def\ISI{ISI}}
\def\IFNA{Interface Foundation of North America (IFNA)
    \gdef\IFNA{IFNA}}

\def\PPDM{privacy-preserving data mining (PPDM)
    \gdef\PPDM{PPDM}}

\def\ITSEW{International Total Survey Error Workshops (ITSEW)
    \gdef\ITSEW{ITSEW}}

\def\STEM{science, technology, engineering and mathematics (STEM)
    \gdef\STEM{STEM}}

\def\ASG{Art and Science Group (A\&SG)
    \gdef\ASG{A\&SG}}

\def\HSLS{High School Longitudinal Study (HSLS:09)
    \gdef\HSLS{HSLS:09}}
\def\NHIS{National Health Interview Survey (NHIS)
    \gdef\NHIS{NHIS}}
\def\FRG{Focused Research Groups in the Mathematical Sciences (FRG)
    \gdef\FRG{FRG}}

\def\IES{Institute of Education Sciences (IES)
    \gdef\IES{IES}}
\def\IOJ{Institute of Justice (IOJ)
    \gdef\IOJ{IOJ}}
\def\EAA{experimental analysis of algorithms (EAA)
    \gdef\EAA{EAA}\gdef\EAAC{EAA}}
\def\EAAC{Experimental analysis of algorithms (EAA)
    \gdef\EAA{EAA}\gdef\EAAC{EAA}}

\def\JPC{\textit{Journal of Privacy and Confidentialty} (JPC)
    \gdef\JPC{\textit{JPC}}}

\def\NWG{NISS Working Group (NWG)
    \gdef\NWG{NWG}\gdef\NWGS{NWGS}}
\def\NWGS{NISS Working Groups (NWGs)
    \gdef\NWGS{NWGS}\gdef\NWG{NWG}}

\def\DHHS{Department of Health and Human Services (DHHS)
    \gdef\DHHS{DHHS}}
\def\OCC{Office of the Comptroller of the Currency (OCC)
    \gdef\OCC{OCC}}

\def\FIPSE{Fund for the Improvement of Postsecondary Education
(FIPSE)
    \gdef\FIPSE{FIPSE}}

\def\FHWA{Federal Highway Administration (FHWA)\gdef\FHWA{FHWA}}

\def\SPAIG{Statistical Partnerships among Academia, Industry and
    Government (SPAIG)\gdef\SPAIG{SPAIG}}
\def\COPSS{Committee of Presidents of Statistical Societies (COPSS)
    \gdef\COPSS{COPSS}}

\def\SDDS{School District Demographics System (SDDS)
    \gdef\SDDS{SDDS}}

\def\ITRE{Institute for Transportation Research and Education (ITRE)
    \gdef\ITRE{ITRE}}
\def\TRB{Transportation Research Board (TRB)
    \gdef\TRB{TRB}}

\def\DTRA{Defense Threat Reduction Agency (DTRA)
    \def\DTRA{DTRA}}

\def\CPTAC{Clinical Proteomic Technology Assessment for Cancer (CPTAC)
    \gdef\CPTAC{CPTAC}}

\def\SRS{Division of Science Resources Statistics (SRS)
    \gdef\SRS{SRS}}
\def\SED{Survey of Earned Doctorates (SED)
    \gdef\SED{SED}}
\def\SDR{Survey of Doctorate Recipients (SDR)
    \gdef\SDR{SDR}}
\def\GSS{Survey of Graduate Students and Postdoctorates in
    Science and Engineering (GSS)
    \gdef\GSS{GSS}}
\def\RCG{National Survey of Recent College Graduates (RCG)
    \gdef\RCG{RCG}}
\def\NSCG{National Survey of College Graduates (NSCG)
    \gdef\NSCG{NCSG}}
\def\BRDIS{Business R\&D Innovation Survey (BRDIS)
    \gdef\BRDIS{BRDIS}}
\def\SANDE{science and engineering (S\&E)
    \gdef\SANDE{S\&E}}
\def\IPEDS{Integrated Postsecondary Education Data System (IPEDS)
    \gdef\IPEDS{IPEDS}}

\def\SEW{science, engineering and health workforce (SEHW)
    \gdef\SEW{SEHW}}

\def\CIPSEA{Confidential Information Protection and Statistical
    Efficiency Act of 2002 (CIPSEA)
    \gdef\CIPSEA{CIPSEA}}

\def\PSU{primary sampling unit (PSU)
    \gdef\PSU{PSU}\gdef\PSUS{PSUs}}
\def\PSUS{primary sampling units (PSUs)
    \gdef\PSU{PSU}\gdef\PSUS{PSUs}}

\def\NHANES{National Health and Nutrition Examination Survey (NHANES)
    \gdef\NHANES{NHANES}}
\def\ECLS{Early Childhood Longitudinal Study (ECLS)
    \gdef\ECLS{ECLS}}

\def\FERPA{Family Educational Rights and Privacy Act (FERPA)
    \gdef\FERPA{FERPA}}

\def\LBD{Longitudinal Business Database (LBD)
    \gdef\LBD{LBD}}
\def\LEHD{Longitudinal Employer-Household Dynamics (LEHD)
    \gdef\LEHD{LEHD}}
\def\SLDS{statewide longitudinal data systems (SLDS)
    \gdef\SLDS{SLDS}}

\def\ECLSK{Early Childhood Longitudinal Study--Kindergarten Class of
    1998--99 (ECLS-K)\gdef\ECLSK{ECLS-K}}
\def\SESTAT{Scientists and Engineers Statistical Data System (SESTAT)
    \gdef\SESTAT{SESTAT}}

\def\RENCI{Renaissance Computing Institute (RENCI)
    \gdef\RENCI{RENCI}}

\def\NIA{National Institute on Aging (NIA)
    \gdef\NIA{NIA}}
\def\SCOPE{Statistical Community of Practice and Engagement (SCOPE)
    \gdef\SCOPE{SCOPE}}
\def\ICSP{Interagency Council on Statistical Policy (ICSP)
    \gdef\ICSP{ICSP}}

\def\WSSM{World's Simplest Survey Microsimulator (WSSM)
    \gdef\WSSM{WSSM}}
\def\CES{Consumer Expenditure Survey (CES)
    \gdef\CES{CES}}

\def\NCSES{National Center for Science and Engineering Statistics (NCSES)
    \gdef\NCSES{NCSES}}

\def\TEP{Technical Expert Panel (TEP)
    \gdef\TEP{TEP}}
\def\ESSIN{Education Statistics Support Institute Network (ESSIN)
    \gdef\ESSIN{ESSIN}}
\def\FCSM{Federal Committee on Statistical Methodology (FCSM)
    \gdef\FCSM{FCSM}}
\def\JSM{Joint Statistical Meetings (JSM)
    \gdef\JSM{JSM}}
\def\NRBA{nonresponse bias analysis (NRBA)
    \gdef\NRBA{NRBA}}
\def\PSS{Private School Survey (PSS)
    \gdef\PSS{PSS}}
\def\CWI{comparable wage index (CWI)
    \gdef\CWI{CWI}}

\def\TSE{total survey error (TSE)
    \gdef\TSE{TSE}}

\def\TCRN{Triangle Census Research Network (TCRN)
    \gdef\TCRN{TCRN}}
\def\CER{comparative effectiveness research (CER)
    \gdef\CER{CER}}

\def\PII{personally identifiable information (PII)
    \gdef\PII{PII}}

\def\OES{Occupational Employment Statistics (OES)
    \gdef\OES{OES}}
\def\CPI{Consumer Price Index (CPI)
    \gdef\CPI{CPI}}
\def\CBSA{Core Based Statistical Area (CBSA)
    \gdef\CBSA{CBSA}\gdef\CBSAS{CBSAs}}
\def\CBSAS{Core Based Statistical Areas (CBSAs)
    \gdef\CBSA{CBSA}\gdef\CBSAS{CBSAs}}
\def\PUMA{Public Use Microdata Area (PUMA)
    \gdef\PUMA{PUMA}\gdef\PUMAS{PUMAs}}
\def\PUMAS{Public Use Microdata Areas (PUMAs)
    \gdef\PUMA{PUMA}\gdef\PUMAS{PUMAs}}
\def\PUMS{public use microdata samples (PUMS)
    \gdef\PUMS{PUMS}}

\def\RDC{Research Data Center (RDC)
    \gdef\RDC{RDC}\gdef\RDCS{RDCs}}
\def\RDCS{Research Data Centers (RDCs)
    \gdef\RDC{RDC}\gdef\RDCS{RDCs}}

\def\NCRN{NSF--Census Research Network (NCRN)
    \gdef\NCRN{NCRN}}
\def\NCRNCO{NSF--Census Research Network Coordination Office (NCRN-CO)
    \gdef\NCRNCO{NCRN-CO}}

\def\MPR{Mathematica Policy Research (MPR)
    \gdef\MPR{MPR}}
\def\NORC{NORC at the University of Chicago (NORC)
    \gdef\NORC{NORC}}
\def\AAPOR{American Association for Public Opinion Research (AAPOR)
    \gdef\AAPOR{AAPOR}}
\def\AEA{American Economic Association (AEA)
    \gdef\AEA{AEA}}

\def\SC{Steering Committee (SC)
    \gdef\SC{SC}}
\def\AEA{American Economic Association (AEA)
    \gdef\AEA{AEA}}
\def\ICES{International Conference on Establishment Statistics (ICES)
    \def\ICES{ICES}}
\def\IRS{Internal Revenue Service (IRS)
    \gdef\IRS{IRS}}
\def\SSA{Social Security Administration (SSA)
    \gdef\SSA{SSA}}
\def\COSSA{Consortium of Social Science Associations (COSSA)
    \gdef\COSSA{COSSA}}
\def\COPAFS{Council of Professional Associations on Federal Statistics (COPAFS)
    \gdef\COPAFS{COPAFS}}
\def\BJS{Bureau of Justice Statistics (BJS)
    \gdef\BJS{BJS}}
\def\ERS{Economic Research Service (ERS)
    \gdef\ERS{ERS}}
\def\USDA{U.S. Department of Agriculture (USDA)
    \gdef\USDA{USDA}}
\def\BEA{Bureau of Economic Analysis (BEA)
    \gdef\BEA{BEA}}

\def\NSO{national statistical office (NSO)
    \gdef\NSO{NSO}\gdef\NSOS{NSOs}}
\def\NSOS{national statistical offices (NSOs)
    \gdef\NSO{NSO}\gdef\NSOS{NSOs}}

\def\PI{Principal Investigator (PI)
    \def\PI{PI}}
\def\PIs{Principal Investigators (PIs)
    \def\PIs{PIs}}
\def\LEHD{Longitudinal Employer-Household Dynamics (LEHD)
    \def\LEHD{LEHD}}
\def\INSEE{Institut national de la statistique et des \'etudes \'economiques (INSEE)
    \def\INSEE{INSEE}}
\def\CREST{Centre de Recherche en \'Economie et Statistique (CREST)
    \def\CREST{CREST}}
\def\IAB{Institut f\"ur Arbeitsmarkt- und Berufsforschung (IAB)
     \def\IAB{IAB}}
\def\CBS{Centraal Bureau voor de Statistiek (CBS)
      \def\CBS{CBS}}
\def\MCU{multipoint conferencing units (MCU)
      \def\MCU{MCU}}

\def\CESCB{Center for Economic Studies (CES)
    \gdef\CESCB{CES}}

\def\SACNAS{Society for Advancement of Chicanos and Native Americans in Science (SACNAS)
    \gdef\SACNAS{SACNAS}}
\def\AWM{Association for Women in Mathematics (AWM)
    \gdef\AWM{AWM}}

\def\USPS{United States Postal Service (USPS)
    \gdef\USPS{USPS}}

\def\HSB{High School and Beyond (HS\&B:82)
    \gdef\HSB{HS\&B:82}}
\def\NELS{National Educational Longitudinal Study (NELS:88)
    \gdef\NELS{NELS:88}}
\def\NLS{National Longitudinal Study (NLS:72)
    \gdef\NLS{NLS:72}}
\def\NDI{National Death Index (NDI)
    \gdef\NDI{NDI}}

\def\NAS{National Academy of Sciences (NAS)
    \gdef\NAS{NAS}}
\def\EPRI{Electric Power Research Institute (EPRI)
    \gdef\EPRI{EPRI}}
\def\PCORI{Patient-Centered Outcomes Research Institute (PCORI)
    \gdef\PCORI{PCORI}}
\def\SHRP2{Strategic Highway Research Program 2 (SHRP 2)
    \def\SHRP2{SHRP 2}}
\def\OMOP{Observational Medical Outcomes Partnership (OMOP)
    \gdef\OMOP{OMOP}}
\def\ASM{Annual Survey of Manufactures (ASM)
    \gdef\ASM{ASM}}
\def\SIPP{Survey of Income and Program Participation (SIPP)
    \gdef\SIPP{SIPP}}

\def\NSRCG{National Survey of Recent College Graduates (NSRCG)
	\gdef\NSRCG{NSCRG}}
\def\SEH{science, engineering and health (SEH)
    \gdef\SEH{SEH}}
\def\HRS{University of Michigan Health and Retirement Study (HRS)
    \gdef\HRS{HRS}}
\def\NPSAS{National Postsecondary Student Aid Study (NPSAS)
    \gdef\NPSAS{NPSAS}}
\def\BnB{Baccalaureate and Beyond (B\&B)
    \gdef\BnB{B\&B}}
\def\BPS{Beginning Postsecondary Students Longitudinal Study (BPS)
    \gdef\BPS{BPS}}

\def\CMSS{Computation Methods in the Social Sciences (CMSS)
    \gdef\CMSS{CMSS}}

\def\TCS{Teacher Compensation Survey (TCS)
	\gdef\TCS{TCS}}
\def\SASS{Schools and Staffing Survey (SASS)
	\gdef\SASS{SASS}}
\def\NEA{National Education Association (NEA)
	\gdef\NEA{NEA}}
\def\FARS{Fatality Analysis and Reporting System (FARS)
	\gdef\FARS{FARS}}
\def\EDA{exploratory data analysis (EDA)
	\gdef\EDA{EDA}}

\def\CAT{computerized adaptive testing (CAT)
    \gdef\CAT{CAT}}
\def\CBT{computer-based testing (CBT)
    \gdef\CBT{CBT}}

\def\IRT{item response theory (IRT)
    \gdef\IRT{IRT}}

\def\CAPI{computer-assisted personal interview (CAPI)
    \gdef\CAPI{CAPI}}
\def\CATI{computer-assisted telephone interview (CATI)
    \gdef\CATI{CATI}}

\def\CAR{conditional autoregressive (CAR)
    \gdef\CAR{CAR}}

\def\IPUMS{Integrated Public Use Microdata Series (IPUMS)
    \gdef\IPUMS{IPUMS}}

\def\LEA{local education authority (LEA)
    \gdef\LEA{LEA}\gdef\LEAS{LEAs}}
\def\SEA{state education authority (SEA)
    \gdef\SEA{SEA}\gdef\SEAS{SEAs}}
\def\LEAS{local education authorities (LEAs)
    \gdef\LEA{LEA}\gdef\LEAS{LEAs}}
\def\SEAS{state education authorities (SEAs)
    \gdef\SEA{SEA}\gdef\SEAS{SEAs}}

\def\CMS{Centers for Medicare and Medicaid Services (CMS)
    \gdef\CMS{CMS}}

\def\MMS{Methodology, Measurement, and Statistics (MMS)
    \gdef\MMS{MMS}}

\def\SOC{standard occupational code (SOC)
    \gdef\SOC{SOC}}
\def\ASCO{American Society for Clinical Oncology (ASCO)
    \gdef\ASCO{ASCO}}

\def\UI{unemployment insurance (UI)
    \gdef\UI{UI}}

\def\PT{Project TALENT (PT)
    \gdef\PT{PT}}

\def\AERA{American Educational Research Association (AERA)
    \gdef\AERA{AERA}}

\def\CBSA{Core Business Statistical Area (CBSA)
    \gdef\CBSA{CBSA}\def\CBSAs{CBSAs}}
\def\CBSAs{Core Business Statistical Areas (CBSAs)
    \gdef\CBSA{CBSA}\def\CBSAs{CBSAs}}

\def\PT{Project Talent (PT)
    \gdef\PT{PT}}
\def\FCB{First Citizens Bank (FCB)
    \gdef\FCB{FCB}}
\def\BWF{Burroughs Wellcome Fund (BWF)
    \gdef\BWF{BWF}}

\def\PPRL{privacy-preserving record linkage (PPRL)
    \gdef\PPRL{PPRL}}

\def\NIA{National Institute on Aging (NIA)
    \gdef\NIA{NIA}}

\def\CHAID{Chi-squared automatic interaction detection (CHAID)
    \gdef\CHAID{CHAID}}
\def\GSC{General School Characteristics (GSC)
    \gdef\GSC{GSC}}

\def\CoDA{Center of Excellence for Complex Data Analysis (CoDA)
    \gdef\CoDA{CoDA}}

\def\AIC{Akaike information criterion (AIC)
    \gdef\AIC{AIC}}
\def\BIC{Bayes information criterion (BIC)
    \gdef\BIC{BIC}}
\def\SSES{Social, Statistical, and Environmental Sciences (SSES)
    \gdef\SSES{SSES}}
\def\SCSS{Survey, Computing, and Statistical Sciences (SCSS)
    \gdef\SCSS{SCSS}}
\def\DSDS{Division for Statistical and Data Science (DSDS)
    \gdef\DSDS{DSDS}}

\def\CM{Census of Manufactures (CM)
    \gdef\CM{CM}}

\def\FSRDC{Federal Statistical Research Data Center (FSRDC)
    \gdef\FSRDC{FSRDC}\gdef\FSRDCS{FSRDCs}}
\def\FSRDCS{Federal Statistical Research Data Centers (FSRDCs)
    \gdef\FSRDC{FSRDC}\gdef\FSRDCS{FSRDCs}}

\def\PTTP{partially trusted third party (PTTP)
    \gdef\PTTP{PTTP}\gdef\PTTPS{PTTPs}}
\def\PTTPS{partially trusted third parties (PTTPs)
    \gdef\PTTP{PTTP}\gdef\PTTPS{PTTPs}}

\def\NAICS{North American Industrial Classification System (NAICS)
    \gdef\NAICS{NAICS}}
\def\CART{classification and regression trees (CART)
    \gdef\CART{CART}}

\def\FEBRL{Freely Extensible Biomedical Record Linkage (FEBRL)
    \gdef\FEBRL{FEBRL}}
\def\FRIL{Fine-Grained Records Integration and Linkage Tool (FRIL)
    \gdef\FRIL{FRIL}}
\def\SOEMPI{Secure Open Enterprise Master Patient Index (SOEMPI)
    \gdef\SOEMPI{SOEMPI}}

\def\SDA{statistical disclosure avoidance (SDA)
    \gdef\SDA{SDA}}

\def\LAS{Laboratory for Analytic Sciences (LAS)
    \gdef\LAS{LAS}}

\def\IPF{iterative proportional fitting (IPF)
    \gdef\IPF{IPF}}
\def\ROC{receiver operating characteristic (ROC)
    \gdef\ROC{ROC}}

\def\NCBI{National Center for Biotechnology Information (NCBI)
    \gdef\NCBI{NCBI}}

\def\SCC{Specified Certainty Classification (SSC)
    \gdef\SCC{SCC}}

\def\BP{base pairs (BP) 
    \gdef\BP{BP}}

\def\ROC{receiver operating characteristic (ROC)
    \gdef\ROC{ROC}}

\gdef\MS{\texttt{Mason\_simulator}}

\gdef\MV{\texttt{Mason\_variator}}

\def\AI{artificial intelligence (AI)
   \gdef\AI{AI}}

\def\SARS{severe acute respiratory syndrome (SARS)
  \gdef\SARS{SARS}}

\def\SNP{single nucleotide polymorphism (SNP)
  \gdef\SNP{SNP}
  \gdef\SNPs{SNPs}}

\def\SNPs{single nucleotide polymorphisms (SNPs)
  \gdef\SNPs{SNPs}
  \gdef\SNP{SNP}}

\def\CRISPR{clustered regularly interspaced short palindromic repeats (CRISPR)
  \gdef\CRISPR{CRISPR}}

\def\CAS{CRISPR associated sequence (CAS)
  \gdef\CAS{CAS}}

\def\MDS{multidimensional scaling (MDS)
  \gdef\MDS{MDS}}

\gdef\MS{\texttt{Mason\_simulator}}

\gdef\MV{\texttt{Mason\_variator}}

\begin{document}

\title{Application of Markov Structure of Genomes to\\
Outlier Identification and Read Classification}

\author{Alan F. Karr\thanks{Fraunhofer USA CMA; akarr@fraunhofer.org; Corresponding author}, Jason Hauzel\thanks{Fraunhofer USA CMA; jhauzel@fraunhofer.org}, Adam A. Porter\thanks{University of Maryland, Department of Computer Science; aporter@cs.umd.edu}, Marcel Schaefer\thanks{Fraunhofer USA CMA; mschaefer@fraunhofer.org}}

\maketitle

\begin{abstract}
In this paper we apply the structure of genomes as second-order Markov processes specified by the distributions of successive triplets of bases to two bioinformatics problems: identification of outliers in genome databases and read classification in metagenomics, using real coronavirus and adenovirus data.
\end{abstract}

\section{Introduction}
That the sequential structure of genomes is important has been known since the discovery of DNA. In this paper we employ a statistics and stochastic process perspective on triplets of successive bases to address two important applications: identifying outliers in genome databases, and classifying reads in the metagenomic context of reference-guided assembly. From this stochastic process perspective, triplets are a second-order Markov chain specified by the distribution of each base conditional on its two immediate predecessors.

To be sure, studying genomes via base sequence distributions is not novel. Previous papers have addressed genome signatures \citep{karlin-biases-1997, campbell-comparisons-1999, takahashi-signatures-2003}, as well as frequentist \citep{rosen-fragclass-2008} and Bayesian \citep{wang-bayesclass-2007} approaches to classification problems. While we focus instead on triplet distributions, many extant works employ quartets, also referred to as tetranucleotides, among them \cite{pride-tetra-2003}, \cite{teeling-tetra-classification-2004} and \cite{teeling-tetra-web-2004}.

In \S \ref{sec.outliers}, we apply hierarchical cluster analysis of triplet distributions to identification of outliers in genome databases. Genomes in small or highly divergent clusters are statistical outliers. They may be data quality problems, as some in our example clearly are, or simply aberrant. The clustering is both statistically effective and scientifically interpretable.

The second application, in \S\ref{sec.readclass}, is Bayesian classification of (short) reads in metagenomics. The latter has consequences for uncertainty quantification, in particular allowing assessment of the contributions to uncertainty of prior knowledge, input data quality, and training data quality, which is also interpretable as model quality. Throughout, we use real data, primarily a coronavirus dataset downloaded from the \NCBI, as well as a single adenovirus genome included in the download package for the \texttt{Art} read simulator \citep{art-2012}. Each genome has length on the order of 30,000 \BP, representing approximately 20 genes. Conclusions appear in \S\ref{sec.conclusion}.

\section{Background}\label{sec.triplets}
In this paper, a genome $G$ is a character string chosen from the base (nucleotide) alphabet $\mathcal{B} = \{A, C, G, T\}$, and represents one strand of the DNA in an organism. Given a genome $G$, the length of $G$ is $|G|$; the $i^{\mathrm{th}}$ base in $G$ is $G(i)$; and the bases from location $i$ to location $j > i$ are $G(i:j)$. For $n \geq 1$, let $\mathcal{B}(n)$ be the set of all sequences of length $n$ from $\mathcal{B}$, of which there are $4^n$.  We focus on triplets, whose distributions are 64-dimensional summaries of genomes;\footnote{Other cases of interest appear in the literature. Base distributions ($n=1$) and pair distributions ($n=2$) are too coarse to be useful for the problems we address. Quartets ($n=4$) have historical precedent
\citep{pride-tetra-2003, teeling-tetra-classification-2004, teeling-tetra-web-2004}. Based on informal experiments, for the applications in \S\ref{sec.outliers} and \ref{sec.readclass}, quartets are more cumbersome computationally than triplets, but do not seem to be significantly more informative.} for this focal case, $\mathcal{B}(3) = \{AAA, AAC, \dots, TTG, TTT\}$, and the triplet distribution $P_3(\cdot|G)$ is defined as
\begin{equation}
P_3(b_1b_2b_3|G) = \mathrm{Prob}\big\{G(k:k+2) = b_1b_2b_3\big\}
\label{eq.tripletdistribution}
\end{equation}
for each choice of $b_1, b_2, b_3$ from $\mathcal{B}$, where $k$ is chosen at random from $1, \dots, |G|-2$. An equivalent perspective is that of a second-order Markov chain. The information contained in $P_3(\cdot|G)$ is the same as that contained in the pair distribution $P_2(\cdot|G)$ and the $16 \times 4$ transition matrix
\begin{equation}
T_3((b_1, b_2), b_3|G) = \mathrm{Prob}\big\{G(k+2) = b_3| G(k) = b_1, G(k+1) = b_2\big\},
\label{eq.transitionmatrix3}
\end{equation}
whose rows are indexed by $(b_1,b_2)$ and columns are indexed by $b_3$, and which gives the distribution of each base conditional on its two predecessors.

Appendix \ref{app.example} contains examples of both distributions for the three virus genomes underlying the read classification problem in \S\ref{sec.readclass}.

\section{Outlier Detection}\label{sec.outliers}
In this section, we use triplet and amino acid distributions to detect outliers in the \NCBI\ database. Outliers may be true scientific anomalies, or they may be data quality problems \citep{dqdegradation-2021}. That large, uncurated problem genome databases may have quality issues is of increasing concern in a variety of contexts \citep{wildpatterns-2018, steinegger-2020, biodefense-2019}.

Our strategy is to use hierarchical cluster analysis \citep{aggarwal-reddy-clustering-2013, everitt-clustering-2011} of the 64-dimensional triplet distributions, and then to investigate clusters with small counts, which \emph{are} outliers in a statistical sense, to see if they are outliers in other senses. As our experimental platform, we employ a coronavirus database containing 26,953 genomes, which was downloaded from the \NCBI\ in November of 2020. To it, we added eleven ''known'' outliers: one adenovirus genome and low-quality versions of ten coronavirus genomes selected randomly from the 26,953. The latter were produced using the degradation methodology described in \cite{dqdegradation-2021}, which consists of repeated iterations of the \MV\ haplotype generation software \citep{fu_mi_publications962}. The clustering was performed on standardized versions of the triplet distributions using the ``complete'' method for hierarchical clustering \citep{fastcluster-2013} in the \texttt{R} software system \citep{R-2021}.

We emphasize that some of what follows requires knowledge of the sources of the genomes (original coronavirus dataset, adenovirus or degraded) that is not available in some other contexts.

\subsection{Clustering Based on Triplet Distributions}\label{subsec.outliers-triplets}
\begin{figure}[htbp]
\begin{center}
\includegraphics[width=4in]{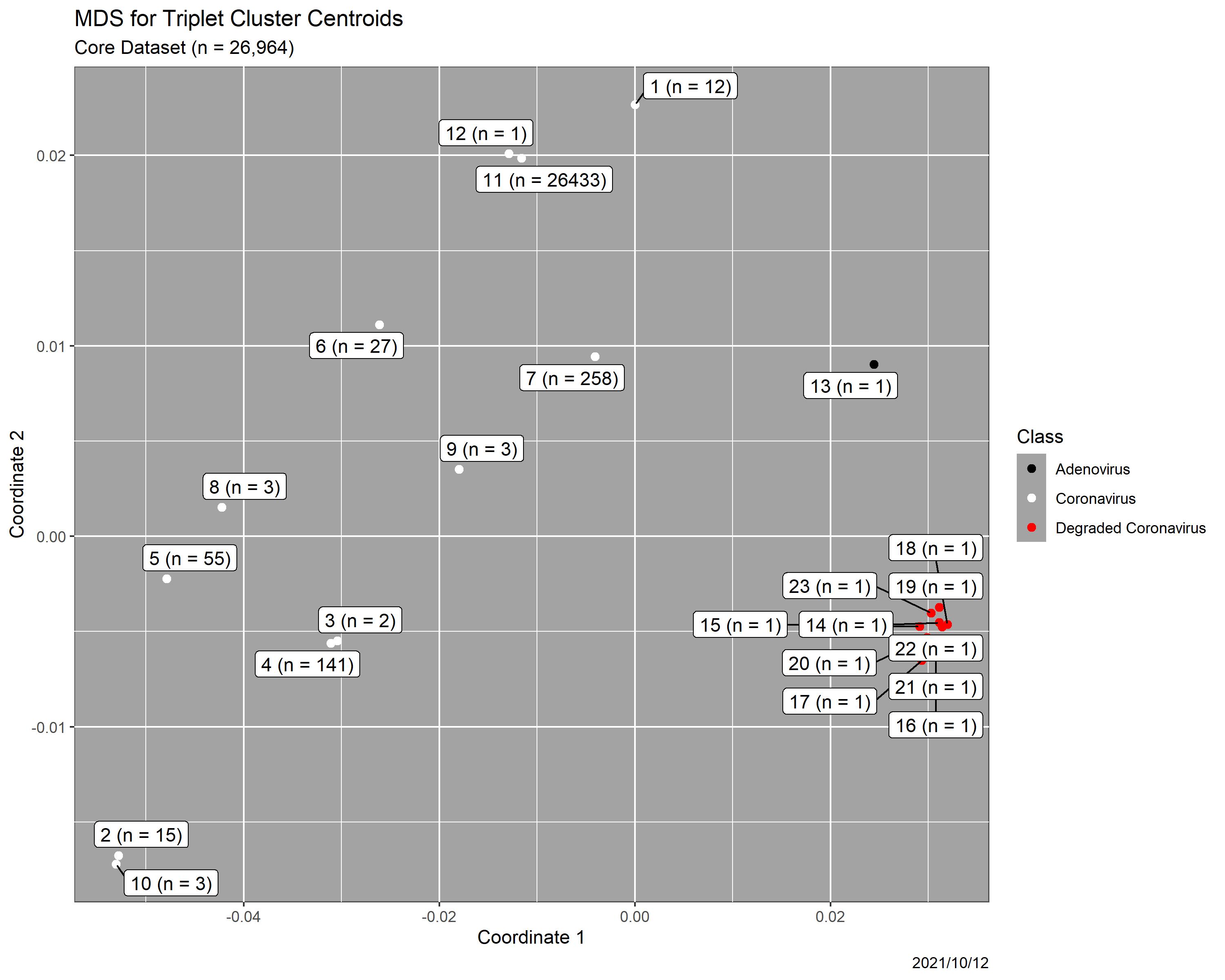}

\vspace{.1in}
\includegraphics[width=5in]{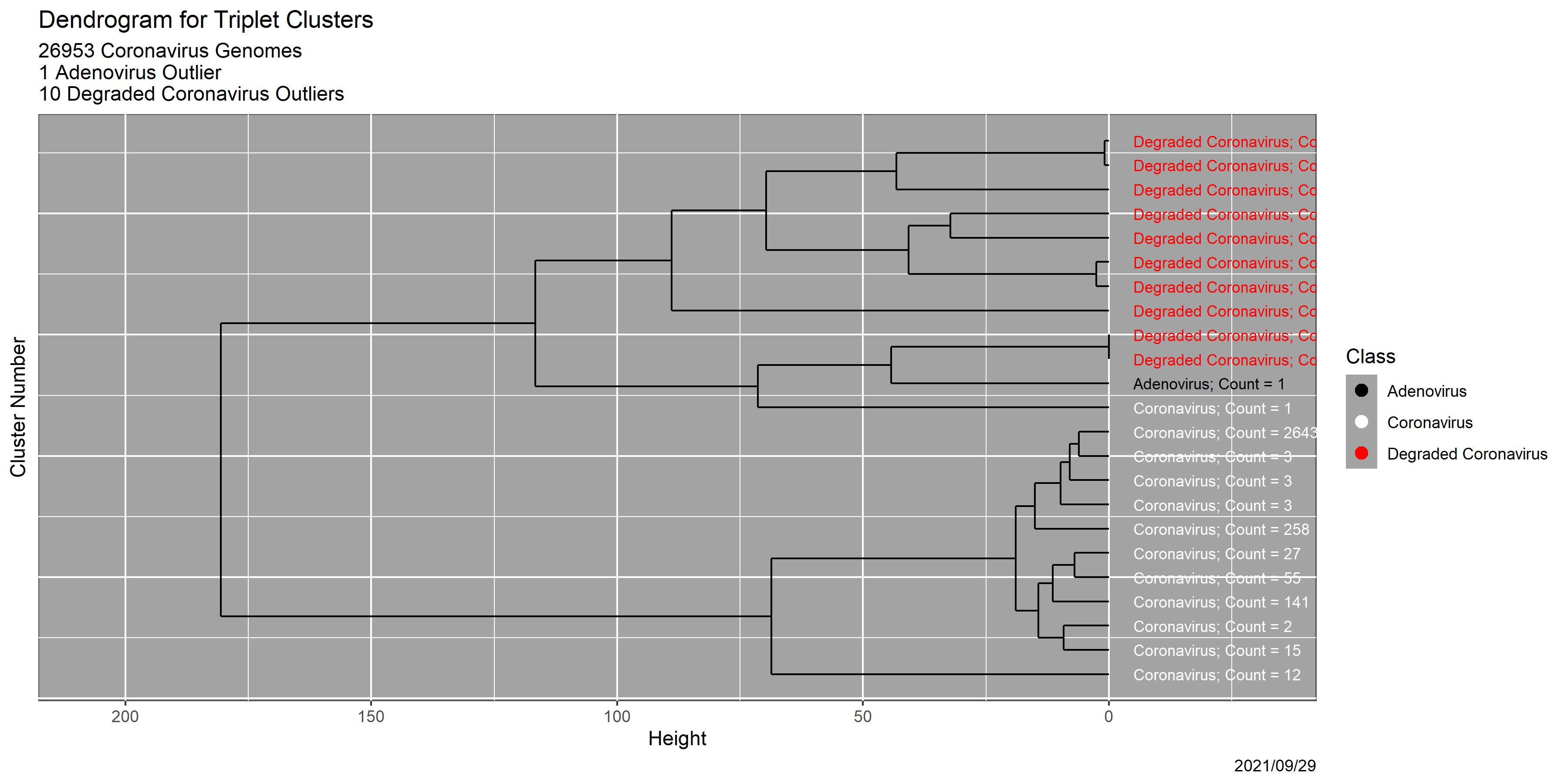}
\end{center}
\caption{\emph{Top:} Two-dimensional MDS plot of triplet distribution of cluster centroids for the 26,964-element coronavirus dataset containing 26,953 coronavirus genomes and 11 deliberately added outliers. \emph{Bottom:} Corresponding dendrogram. Class in both plots is the same as in Table \ref{tab.clusterpropeties-triplets}, and is shown in the same colors in both graphics.}
\label{fig.clustering}
\end{figure}

Figure \ref{fig.clustering} contains results for triplet-based hierarchical clustering in 64 dimensions.\footnote{Technically, in a 63-dimensional hyperplane, since the 64 probabilities sum to one.} The number of clusters, which is 23, was determined using standard heuristics that trade off complexity for explanatory power. The top panel is a plot of the cluster centroids reduced to two dimensions using \MDS\ \citep{kruskal-mds-1978, cox-cox-mds-2001}. The greater the separation in the plot, the greater the dissimilarity between clusters. The bottom panel of Figure \ref{fig.clustering} is the associated dendrogram, depicting the dynamics of the clustering process, and is discussed in more detail below.

Statistically, the quality of the clustering is very high. The percentage of the variation of each triplet frequency, over the entire 26,964-genome dataset, that is explained by the cluster numbers alone is very large. Specifically, for each triplet $x$, a linear model of the form $P_3(x|\cdot) \sim \mathrm{ClusterNumber}(\cdot)$ was fitted, treating the latter as a categorical variable. All of the coefficients of determination, $R^2$, exceed 0.93, and the majority exceed 0.98.

Table \ref{tab.clusterpropeties-triplets} complements the dendrogram in Figure \ref{fig.clustering}. In the dendrogram, the $x$-axis, \emph{Height}, is the stage is the clustering process at which splits occur; the greater the height, the earlier the split. So, for example, the first split in the dendrogram creates two branches, the upper one containing the eleven added outliers and one coronavirus genome, and the lower containing the remaining 26,952 coronavirus genomes. Although this is not a logical necessity, every final cluster contains only one ``kind'' of genome, labeled \emph{Class}, which appears in the \MDS\ plot, the dendrogram and the table.\footnote{``Kind'' was termed source previously, and is one of coronavirus, adenovirus or degraded coronavirus.} \emph{Separation Height} Table \ref{tab.clusterpropeties-triplets}  is that at which the cluster is separated from all the others and not split further. The higher this value, the earlier in the process a cluster is permanently split off from the others. \emph{Separation Rank} orders the clusters by Separation Height, with the greatest value of height having rank 1. Ties end in .5.

\begin{table}[htbp]
\begin{center}
\begin{tabular}{lrlrr}
  \hline
Cluster & Count & Class & Separation Height & Separation Rank \\
  \hline
  1 &  12 & Coronavirus & 68.60 & 3 \\
  2 &  15 & Coronavirus & 9.23 & 11.5 \\
  3 &   2 & Coronavirus & 9.23 & 11.5 \\
  4 & 141 & Coronavirus & 11.42 & 9 \\
  5 &  55 & Coronavirus & 6.98 & 14.5 \\
  6 &  27 & Coronavirus & 6.98 & 14.5 \\
  7 & 258 & Coronavirus & 15.03 & 8 \\
  8 &   3 & Coronavirus & 9.87 & 10 \\
  9 &   3 & Coronavirus & 7.99 & 13 \\
  10 &   3 & Coronavirus & 6.08 & 16.5 \\
  11 & 26433 & Coronavirus & 6.08 & 16.5 \\
  12 &   1 & Coronavirus & 71.43 & 2 \\ \hline\hline
  13 &   1 & Adenovirus & 44.29 & 4 \\
  14 &   1 & Degraded Coronavirus & 0.00 & 22.5 \\
  15 &   1 & Degraded Coronavirus & 0.00 & 22.5 \\
  16 &   1 & Degraded Coronavirus & 88.92 & 1 \\
  17 &   1 & Degraded Coronavirus & 2.58 & 18.5 \\
  18 &   1 & Degraded Coronavirus & 2.58 & 18.5 \\
  19 &   1 & Degraded Coronavirus & 32.20 & 6.5 \\
  20 &   1 & Degraded Coronavirus & 32.20 & 6.5 \\
  21 &   1 & Degraded Coronavirus & 43.27 & 5 \\
  22 &   1 & Degraded Coronavirus & 0.92 & 20.5 \\
  23 &   1 & Degraded Coronavirus & 0.92 & 20.5 \\
   \hline
\end{tabular}
\caption{Characteristics of the triplet distribution clusters in Figure \ref{fig.clustering}. See text for explanation of Separation Height and Separation Rank. The double horizontal line divides original 26,953 coronavirus genomes from the 11 deliberate outliers.}
\label{tab.clusterpropeties-triplets}
\end{center}
\end{table}

We now interpret. The overwhelming majority of coronavirus genomes---26,433 of the original 26,953, or 98.1\%---are in cluster 11. Cluster 13 contains the adenovirus genome alone; each of the 10 degraded coronavirus genomes appears in a cluster by itself (clusters 14--23). Therefore, the deliberate outliers are not only all detected, but also distinguished from one another.\footnote{We caution against over-interpretation. Some of the behavior is a consequence of the small number of deliberate outliers.} The MDS plot in Figure \ref{fig.clustering} shows that these 10 degraded coronavirus outliers differ more from the coronavirus genomes (clusters 1--12) and the adenovirus (cluster 13) than from one another. The adenovirus genome differs from both the coronavirus genomes and the degraded coronavirus outliers.

One original coronavirus genome appears by itself, in cluster 12. Without question it is an outlier; as noted above, it is separated from the other coronavirus genomes at the first split. However, we see below that when clustering is based on amino acids instead of triplets, it ceases to be an outlier. The \NCBI\ dataset ID of this genome is \emph{MT451283.1 Severe acute respiratory syndrome coronavirus 2 isolate SARS-CoV-2/human/AUS/VIC402/2020, complete genome}. No other genome ID in the database is similar to it, which we take as scientific confirmation that the triplet distribution-based clustering truly detects outliers.

Cluster 16 has Separation Rank 1: that particular degraded coronavirus genome is the first to be separated permanently from everything else. Cluster 12, with the singleton outlier coronavirus genome, is second. Of the remaining clusters (1--10), cluster 1 is the strongest candidate for containing outliers. Not only does it have Separation Rank 3, but also its count is small. The next two highest ranked coronavirus clusters (4 and 7) do not have similarly small counts, but based on the results here and in \S\ref{subsec.outliers-aminoacids} are nevertheless compelling candidates to be outliers.

Returning to the ID-based interpretation of cluster 12, Table \ref{tab.IDsByCluster-triplets} shows that every coronavirus cluster 1--10 (i.e., all except the the outlier constituting cluster 12 and the cluster with 26,433 members) is essentially defined by a substring of the ID, subject to two qualifications. First, some clusters (to wit, 1, 4, and 7) contain multiple substrings that cover the cluster and appear in no other cluster. And second, there are substrings that cover multiple clusters but appear in no other clusters. For instance, ``Human coronavirus HKU1'' covers all 15 elements of cluster 2 and all 3 elements of cluster 10, and appears nowhere else. We again conclude that the clustering captures scientific reality.

\begin{table}[htbp]
\begin{footnotesize}\begin{center}
\begin{tabular}{|l|r|l|r|r|}
\hline
Cluster & Cluster & Substring of ID                             & \multicolumn{2}{c|}{Count in}
\\
        & Count  &                                              & This Cluster & Other Clusters
\\
\hline\hline
1       & 12  & TOTAL                                           &              &
\\
        &     & SARS coronavirus Tor2                           & 2            & 0
\\
        &     & SARS coronavirus GDH-BJH01                      & 1            & 0
\\
        &     & SARS coronavirus P2                             & 1            & 0
\\
        &     & SARS coronavirus HKU-39849 isolate TCVSP-HARROD & 2            & 0
\\
        &     & SARS coronavirus Urbani isolate icSARS          & 6            & 0
\\
\hline
2       & 15  & Human coronavirus HKU1                          & 15           & 3
\\
10      & 3   &                                                 & 3            & 15
\\
\hline
3       & 2   & Human enteric coronavirus [\dots] 4408          & 2            & 0
\\
\hline
4       & 141 & TOTAL                                           &              &
\\
        &     & Human coronavirus OC43 strain                   & 138          & 0
\\
        &     & Coronavirus cy[abc]-BetaCoV/2019                & 3            & 0
\\
\hline
5       & 55  & Human coronavirus NL63 strain                   & 55           & 6
\\
8       & 3   &                                                 & 3            & 58
\\
9       & 3   &                                                 & 3            & 58
\\
\hline
6       & 27  & Human coronavirus 229E                          & 27           & 0
\\
\hline
7       & 258 & TOTAL                                           &              &
\\
        &     & Middle East respiratory syndrome                & 253          & 0
\\
        &    & [Bb]etacoronavirus                               & 5            & 0
\\
\hline
12      & 1   & SARS-CoV-2/human/AUS/VIC402/2020                & 1            & 0
\\
\hline
\end{tabular}
\caption{Substrings of \NCBI\ IDs that characterize the clusters containing coronavirus genomes. Single-member clusters comprised of the adenovirus genome or one of the ten degraded coronavirus genomes are excluded.}
\label{tab.IDsByCluster-triplets}
\end{center}\end{footnotesize}
\end{table}


\subsection{Clustering Based on Amino Acids}\label{subsec.outliers-aminoacids}
\begin{figure}[htbp]
\begin{center}
\includegraphics[width=4in]{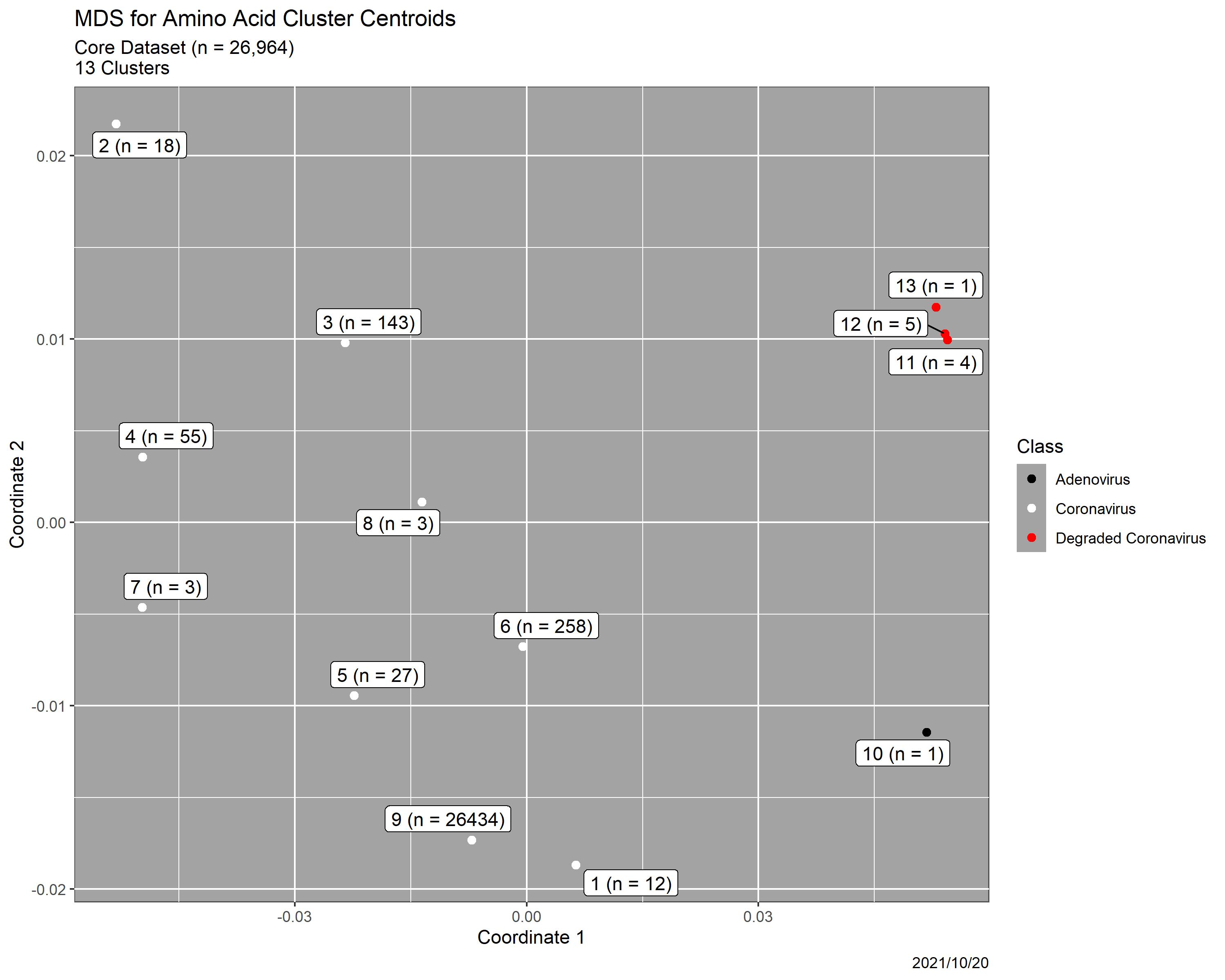}

\vspace{.1in}
\includegraphics[width=5in]{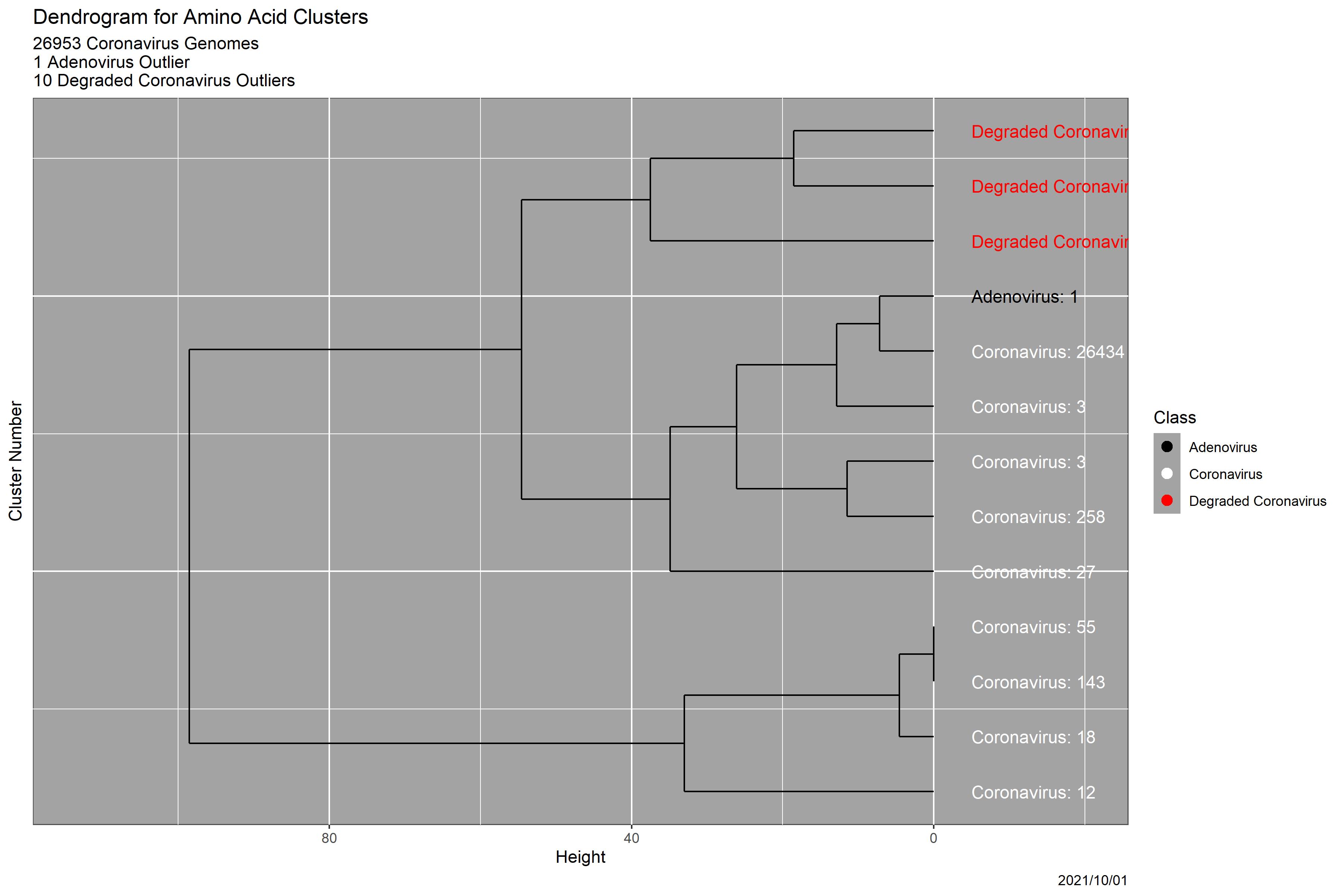}
\end{center}
\caption{\emph{Top:} Two-dimensional MDS plot of amino acid distribution of cluster centroids for the 26,964-element coronavirus dataset containing 26,953 coronavirus genomes and 11 deliberately added outliers.  \emph{Bottom:} Corresponding dendrogram. Class in the dendrogram is the same as in Table \ref{tab.clusterpropeties-AA}, and shown the the same colors in both graphics.}
\label{fig.clustering-AA}
\end{figure}

Clustering was also performed on amino acid distributions,\footnote{Generated directly from triplet distributions using the known many-to-one mapping of triplets to amino acids.} in 21 dimensions rather than 64, resulting in the MDS plot and dendrogram shown in Figure \ref{fig.clustering-AA} and the cluster characteristics in Table \ref{tab.clusterpropeties-AA}. There are fewer clusters than for triplets---13 as compared to 23. The statistical explanatory power matches that for triplets: all coefficients of determination exceed 0.95.

\begin{table}[htbp]
\begin{center}
\begin{tabular}{lrlrr}
  \hline
Cluster & Count & Class & Separation Height & Separation Rank \\
  \hline
  1 & 12 & Coronavirus & 33.0 & 3.0 \\
  2 & 18 & Coronavirus & 4.6 & 11.0 \\
  3 & 143 & Coronavirus & 0.0 & 12.5 \\
  4 & 55 & Coronavirus & 0.0 & 12.5 \\
  5 & 27 & Coronavirus & 34.9 & 2.0 \\
  6 & 258 & Coronavirus & 11.5 & 7.5 \\
  7 &  3 & Coronavirus & 11.5 & 7.5 \\
  8 &  3 & Coronavirus & 12.8 & 6.0 \\
  9 & 26434 & Coronavirus & 7.2 & 9.5 \\ \hline\hline
  10 &  1 & Adenovirus & 7.2 & 9.5 \\
  11 &  4 & Degraded Coronavirus & 37.5 & 1.0 \\
  12 &  5 & Degraded Coronavirus & 18.5 & 4.5 \\
  13 &  1 & Degraded Coronavirus & 18.5 & 4.5 \\
   \hline
\end{tabular}
\caption{Characteristics of the amino acid distribution clusters in Figure \ref{fig.clustering-AA}. See text for explanation of Separation Height and Separation Rank. The double horizontal line divides coronavirus genomes from the deliberate outliers.}
\label{tab.clusterpropeties-AA}
\end{center}
\end{table}

Table \ref{tab.clusters-crosstab}, a cross-tabulation of triplet distribution clusters (rows) and amino acid distribution clusters (columns), shows that each amino acid cluster is either identical to a triplet distribution cluster or an amalgam of triplet distribution clusters. Specifically, for the deliberate outliers, the adenovirus genome remains in a cluster of its own (13 for triplets; 10 for amino acids), while the 10 degraded coronavirus genomes (triplet clusters 14--23) are condensed to three amino acid clusters (11--13). Interestingly, the coronavirus genome that alone comprised triplet cluster 12 is absorbed into the massive cluster (triplet cluster 11), which now has size 26,434 (amino acid cluster 9).
Triplet cluster 1 (count = 12) and amino acid cluster 1 are identical, as are triplet cluster 5 (count = 55) and amino acid cluster 12, as well as triplet cluster 6 (count = 27) and amino acid cluster 5, triplet cluster 7 (count = 258) and amino acid cluster 6, triplet cluster 8 (count = 3) and amino acid cluster 7 and, finally, triple cluster 9 (count = 3) and amino acid cluster 10. Triplet clusters 2 and 10 are merged to create amino acid cluster 2, while triplet clusters 3 and 4 merge into amino acid cluster 3.

\begin{table}[htbp]
\begin{scriptsize}\begin{center}
\begin{scriptsize}
\begin{tabular}{rrrrrrrrrrrrrrr}
\hline
 & \multicolumn{14}{c}{Amino Acid Cluster}
\\
Triplet Cluster & 1 & 2 & 3 & 4 & 5 & 6 & 7 & 8 & 9 & 10 & 11 & 12 & 13 & Sum \\
\hline
1 & \textbf{\large 12} & 0 & 0 & 0 & 0 & 0 & 0 & 0 & 0 & 0 & 0 & 0 & 0 & 12 \\
  2 & 0 & 15 & 0 & 0 & 0 & 0 & 0 & 0 & 0 & 0 & 0 & 0 & 0 & 15 \\
  3 & 0 & 0 & 2 & 0 & 0 & 0 & 0 & 0 & 0 & 0 & 0 & 0 & 0 & 2 \\
  4 & 0 & 0 & 141 & 0 & 0 & 0 & 0 & 0 & 0 & 0 & 0 & 0 & 0 & 141 \\
  5 & 0 & 0 & 0 & \textbf{\large 55} & 0 & 0 & 0 & 0 & 0 & 0 & 0 & 0 & 0 & 55 \\
  6 & 0 & 0 & 0 & 0 & \textbf{\large 27} & 0 & 0 & 0 & 0 & 0 & 0 & 0 & 0 & 27 \\
  7 & 0 & 0 & 0 & 0 & 0 & \textbf{\large 258} & 0 & 0 & 0 & 0 & 0 & 0 & 0 & 258 \\
  8 & 0 & 0 & 0 & 0 & 0 & 0 & \textbf{\large 3} & 0 & 0 & 0 & 0 & 0 & 0 & 3 \\
  9 & 0 & 0 & 0 & 0 & 0 & 0 & 0 & \textbf{\large 3} & 0 & 0 & 0 & 0 & 0 & 3 \\
  10 & 0 & 3 & 0 & 0 & 0 & 0 & 0 & 0 & 0 & 0 & 0 & 0 & 0 & 3 \\
  11 & 0 & 0 & 0 & 0 & 0 & 0 & 0 & 0 & 26433 & 0 & 0 & 0 & 0 & 26433 \\
  12 & 0 & 0 & 0 & 0 & 0 & 0 & 0 & 0 & 1 & 0 & 0 & 0 & 0 & 1 \\
  13 & 0 & 0 & 0 & 0 & 0 & 0 & 0 & 0 & 0 & \textbf{\large 1} & 0 & 0 & 0 & 1 \\
  14 & 0 & 0 & 0 & 0 & 0 & 0 & 0 & 0 & 0 & 0 & 1 & 0 & 0 & 1 \\
  15 & 0 & 0 & 0 & 0 & 0 & 0 & 0 & 0 & 0 & 0 & 0 & 1 & 0 & 1 \\
  16 & 0 & 0 & 0 & 0 & 0 & 0 & 0 & 0 & 0 & 0 & 0 & 0 & \textbf{\large 1} & 1 \\
  17 & 0 & 0 & 0 & 0 & 0 & 0 & 0 & 0 & 0 & 0 & 0 & 1 & 0 & 1 \\
  18 & 0 & 0 & 0 & 0 & 0 & 0 & 0 & 0 & 0 & 0 & 0 & 1 & 0 & 1 \\
  19 & 0 & 0 & 0 & 0 & 0 & 0 & 0 & 0 & 0 & 0 & 0 & 1 & 0 & 1 \\
  20 & 0 & 0 & 0 & 0 & 0 & 0 & 0 & 0 & 0 & 0 & 1 & 0 & 0 & 1 \\
  21 & 0 & 0 & 0 & 0 & 0 & 0 & 0 & 0 & 0 & 0 & 1 & 0 & 0 & 1 \\
  22 & 0 & 0 & 0 & 0 & 0 & 0 & 0 & 0 & 0 & 0 & 1 & 0 & 0 & 1 \\
  23 & 0 & 0 & 0 & 0 & 0 & 0 & 0 & 0 & 0 & 0 & 0 & 1 & 0 & 1 \\
  Sum & 12 & 18 & 143 & 55 & 27 & 258 & 3 & 3 & 26434 & 1 & 4 & 5 & 1 & 26964 \\
\hline
\end{tabular}
\end{scriptsize}
\end{center}
\end{scriptsize}
\caption{Cross-tabulation of triplet-based clusters and amino acid-based clusters. Entries corresponding to identical clusters are highlighted in \textbf{\large enlarged boldface}.}
\label{tab.clusters-crosstab}
\end{table}

The dendrogram in Figure \ref{fig.clustering-AA} and the separation heights and ranks in Table \ref{tab.clusterpropeties-AA} add insightful nuances. In the dendrogram, the first split isolates amino acid clusters 1--4 from everything else, including the massive coronavirus cluster and the deliberate outliers. The next split isolates the ten degraded coronavirus genomes from the latter, but the adenovirus genome remains coupled with the dominant coronavirus cluster until relatively late in the process (height = 7.2), which is the separation height for both. Ignoring the deliberate outliers, there are are two extreme separation heights---34.9 for cluster 5 and 33.0 for cluster 1. The latter reinforces our hypothesis in \S\ref{subsec.outliers-triplets} that cluster 1 contains outliers. The analogous message regarding cluster 5 is attenuated, so using amino acids provides information not available from triplets alone. The very last split in the amino acid dendrogram, which separates cluster 3 (count = 143) and cluster 4 (count = 55) is less discernible in the dendrogram in Figure \ref{fig.clustering}.

\section{Bayesian Classification of Simulated Illumina Reads}\label{sec.readclass}
In this section, we apply triplet distributions to the metagenomics problem of classifying short genome reads in mixtures comprising multiple genomes, a common step in reference-guided assembly.

\subsection{The Experiment}\label{subsec.readclass-experiment}
The components of the experiment are introduced here.

\textbf{Reference Genomes.} There are three reference genomes: the adenovirus genome of length 34,125, downloaded with the read simulator \texttt{Art}, which we also call Adeno; a SARS-CoV-2 genome of length 29,926 contained in the \NCBI\ coronavirus dataset employed in \S \ref{sec.outliers}, which we call COVID; and a \SARS\ genome of length 29,751 from the same database, which we call SARS. Tables \ref{tab.tripletdistributions} and \ref{tab.aminoaciddistributions}, in Appendix \ref{app.example}, show the triplet and amino acid distributions for these three genomes A natural question is ``How different are these triplet distributions?'' Measured by Hellinger distance \citep{nikulin-hd-2010} they are very different. The distances are 0.234 for adenovirus/COVID, 0.125 for adenovirus/SARS and 0.161 for COVID/SARS. To gauge these differences, the empirical 0.001 $p$-values for distributions of genomes of the same size simulated from each distribution are 0.01941755 for adenovirus, 0.02094808 for COVID, and 0.02065539 for SARS. The Hellinger distances for amino acid distributions are 0.189 for adenovirus/COVID, 0.090 for adenovirus/SARS and 0.130 for COVID/SARS. All these distances exceed empirical 0.1\% $p$-values.

\textbf{Read Generation.} The \MS\ read simulator \citep{fu_mi_publications962} was used to simulate Illumina\footnote{Illumina is a major manufacturer of high-end instruments for genome sequencing; their technology is optical in nature; see https://www.illumina.com/.} reads of length 101 from each genome, with approximate 6X coverage. The numbers of reads are 1966, 1996 and 1907, respectively; the total number of reads is 5869. The \MS\ introduces errors in the form of transpositions (\SNPs), insertions, deletions and undetermined bases, which, following convention, appear in the simulated reads as ``N'' and must be accommodated in computation of likelihood functions. Parameters of the \MS\ were set at default values.

\textbf{Likelihood Functions.} Three likelihood functions, one for each genome and denoted by $L_A(\cdot)$, $L_C(\cdot)$ and $L_S(\cdot)$, were calculated, representing the triplet distributions. To illustrate for adenovirus,
\begin{eqnarray}
\lefteqn{L_A(R) = P_2(R(1)R(2)|A) \times T_3((R(1), R(2)), R(3)|A)} \nonumber
\\
& &
\times T_3((R(2), R(3)), R(4)|A) \times \dots \times T_3\bigg(\big(R(|R|-2), R(|R|-1)\big), R(|R|)|A\bigg),
\label{eq.likelihood}
\end{eqnarray}
where $R(k)$ is the $k^{\mbox{th}}$ element of the read $R$, $P_2(\cdot|A)$ is the pair distribution, and $T_3$ is given by (\ref{eq.transitionmatrix3}). In (\ref{eq.likelihood}), we have ignored Ns for simplicity; when they are present, they lead to sums over all possible bases. Below, we also explore alterative likelihood functions representing deliberate degradation of quality of the training data by means of the \MV.

\textbf{Prior and Posterior Probabilities.} For each read $R$, we specify a prior probability distribution $\pi_R$ over the set $\mathcal{A} = \{\mbox{Adeno, COVID, SARS}\}$ of the three genomes, and then use Bayes' theorem and the three likelihoods to calculate posterior probabilities over $\mathcal{A}$, which we then analyze. Specifically, for each read $R$, the posterior probability of $x \in \mathcal{A}$ is given by
\begin{equation}
p(x|R) = \frac{\pi_R(x)L_x(R)}{\pi_R(A)L_A(R) + \pi_R(C)L_C(R) + \pi_R(S)L_S(R)}.
\label{eq.bayes}
\end{equation}

\textbf{Experimental Protocol.} All three components in the calculation can be varied, and their contributions to the final output resolved:
\begin{description}
\item[The input data,]
that is, the reads themselves, which contain \MS-generated errors. Because the \MS\ records the starting location of each read, we have as well the error-free reads, which are of higher quality. Conversely, we can use the \MV\ to degrade read quality, as in \cite{dqdegradation-2021}.
\item[The prior distributions,]
which in most cases we take to be (mildly) informative distributions generated randomly from a Dirichlet distribution. An alternative representing no external prior knowledge is the non-informative priors $\pi_R = (1/3, 1/3, 1/3)$.
\item[The models,]
embedded in the likelihood functions for the three genomes. The three genomes themselves are, in effect, the training data. As discussed further below, model quality can be decreased by degrading them using the \MV.
\end{description}

\subsection{The Base Case}\label{subsec.readclass-base}
\begin{figure}[htbp]
\begin{center}
\includegraphics[width=6.5in]{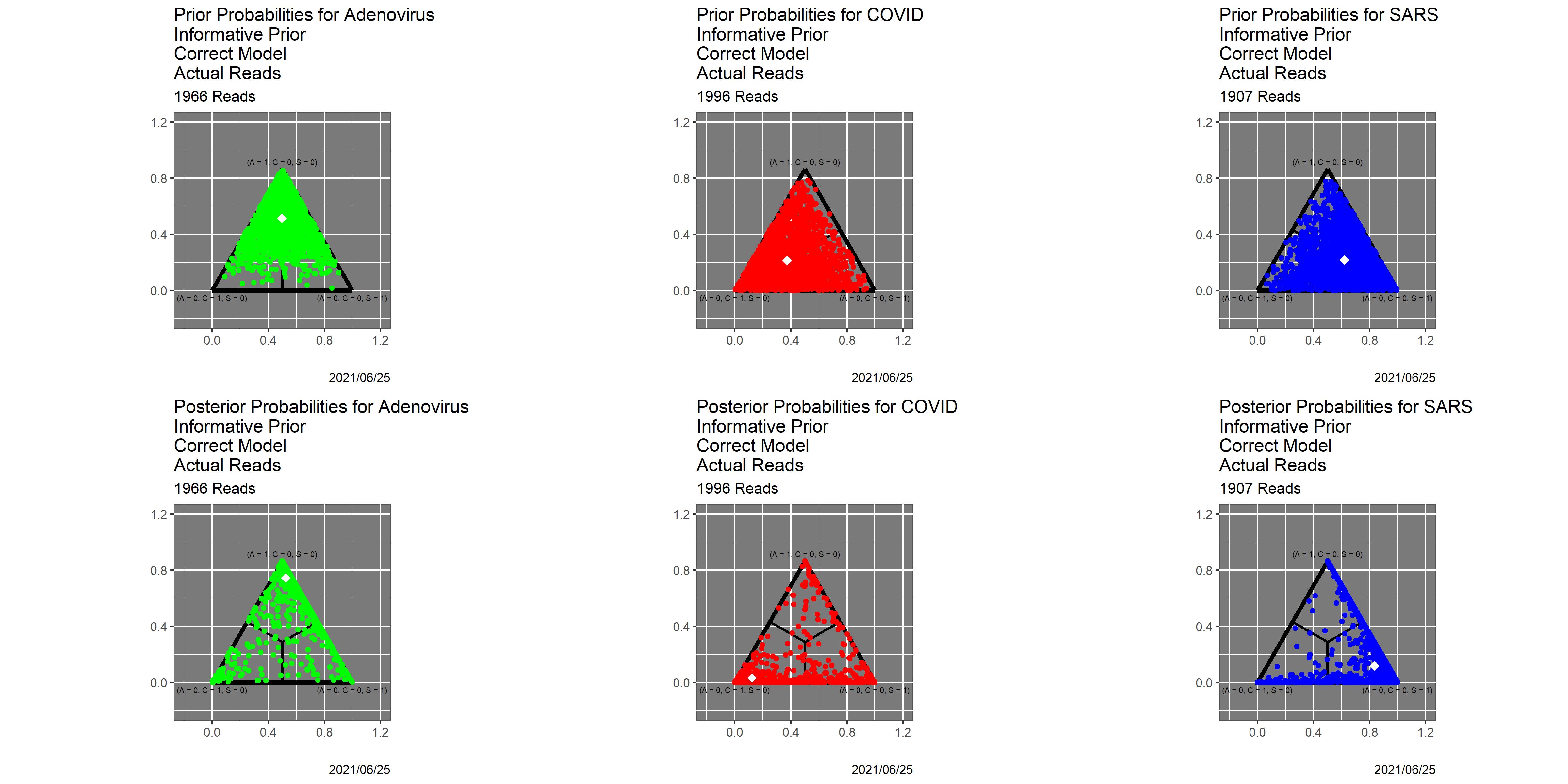}
\end{center}
\caption{Prior and posterior probabilities for the base case of actual reads, informative prior distributions and correct triplet distribution-based models. \emph{Top left:} Prior probabilities for reads from adenovirus. \emph{Top middle:} Prior probabilities for reads from COVID. \emph{Top right:} Prior probabilities for reads from SARS. \emph{Bottom left:} Posterior probabilities for reads from adenovirus. \emph{Bottom middle:} Posterior probabilities for reads from COVID. \emph{Bottom right:} Posterior probabilities for reads from SARS. White symbols in each plot are centroids.}
\label{fig.bayes-basecase}
\end{figure}

As base case, we use the informative priors, the actual reads and the correct models based on the triplet distribution likelihood functions from the three genomes. Figure \ref{fig.bayes-basecase} contains the results. Because similar figures follow, we discuss it in some detail. First, three-dimensional probabilities (barycentric coordinates) are represented in the figure using Cartesian coordinates, as points in a two-dimensional simplex---an equilateral triangle. Pure adenovirus, in the sense that $\mathrm{Prob}(\mbox{Adeno}) = 1$, is the top vertex, pure COVID is the lower left vertex, and pure SARS is the lower right vertex. Because we know the sources of the reads, we create separate displays for each source---adenovirus at the left, COVID in the center and SARS at the right. The upper three panels show prior probabilities, while the lower three panels show posterior probabilities. The green/red/blue coloring depicting read source is redundant, but useful. The white diamond in each scatterplot is the centroid of the probabilities it contains.

The interior black lines, which are often not completely visible, are the decision boundaries for the MAP (maximum \emph{a posteriori} probability) classifier, which assigns to each read the genome with the highest posterior probability. The value of the largest posterior probability cannot be less than 1/3, but can be arbitrarily close to it. Points that lie near any of these boundaries are close calls, but are not distinguished by the MAP classifier from clear-cut cases.\footnote{To note an extreme example, the MAP classifier does not distinguish read $R$ with $p(\cdot|R) = (0.34, 0.33, 0.33)$---an extremely close call---from read $R'$ with $p(\cdot|R') = (0.99, 0.005, 0.005)$---a near-certainty to be from the adenovirus genome. In some contexts, this may be problematic.} Alternative decision strategies in \cite{scc-2021} implement a concept known as \SCC\ that enforces a user-specified lower bound on the certainty of each decision.

Now to interpretations. That the prior distributions are informative is manifested in that each, while scattered, gravitates toward the vertex representing the truth. From the prior to the posterior, all probabilities shift in the direction of the truth--which is what data are supposed to do. Centroids move in response. We caution that there is massive overplotting. For instance, the posterior probabilities for adenovirus (lower left) appear scattered, yet the centroid is very close to the truth---the upper vertex. Table \ref{tab.confusion-basecase} shows the base case confusion matrix for the MAP classifier. The performance is imperfect but respectable: the correct classification rate is $(1757 + 1762 + 1549)/5869 = 86.35$\%.

\begin{table}[htbp]
\begin{center}
\begin{tabular}{|l|r|r|r|r|}
\hline
 & \multicolumn{3}{c|}{Decision} &
\\
\hline
Source & Adeno & COVID & SARS & Sum
\\
\hline
Adeno & 1757  &  74 & 135 & 1966
\\
\hline
COVID &   64 & 1762 & 170 & 1996
\\
\hline
SARS  &  214 &  144 & 1549 & 1907
\\
\hline
Sum  &  2035 & 1980 & 1854 & 5869
\\
\hline
\end{tabular}
\end{center}
\caption{Confusion matrix for the MAP classifier for the base case. Rows are read sources; columns are MAP classifier decisions. The correct classification rate is 86.35\%.}
\label{tab.confusion-basecase}
\end{table}

\subsection{Variations on the Base Case}\label{subsec.readclass-variations}
In terms of understanding all the decisions the classifier \emph{can make}, as opposed to the decisions it \emph{does make} in a specific application \citep{recordlinkage-plos1-2019}, the Bayesian framework allows investigation and quantification of the relative contributions to uncertainty of the priors, the data and the models. Here we explore the effect of each.

\begin{figure}[htbp]
\begin{center}
\includegraphics[width=6.5in]{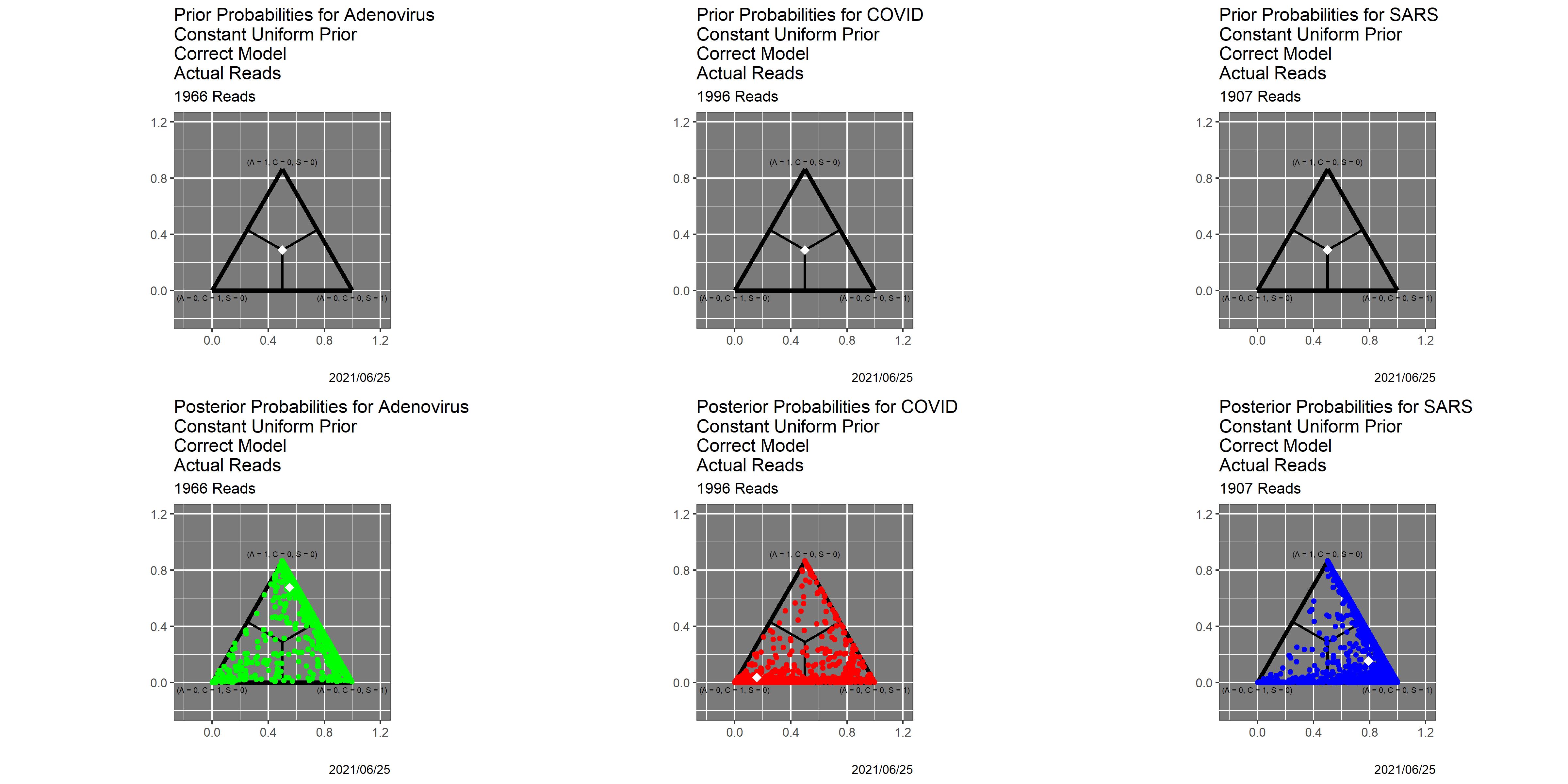}
\end{center}
\caption{Prior and posterior probabilities for the case of actual reads, ``no knowledge priors'' and correct triplet distribution-based models. \emph{Top left:} Prior probabilities for reads from adenovirus. \emph{Top middle:} Prior probabilities for reads from COVID. \emph{Top right:} Prior probabilities for reads from SARS. \emph{Bottom left:} Posterior probabilities for reads from adenovirus. \emph{Bottom middle:} Posterior probabilities for reads from COVID. \emph{Bottom right:} Posterior probabilities for reads from SARS. White symbols in each plot are centroids.}
\label{fig.bayes-uniformpriorbasecase}
\end{figure}

\textbf{Changing the Priors.} The priors are relatively the least important of the three, with one important \emph{caveat} discussed momentarily.. Figure \ref{fig.bayes-uniformpriorbasecase} is the analog of Figure \ref{fig.bayes-basecase} for the ``no prior knowledge'' case that $\pi_R = (1/3, 1/3, 1/3)$ for all reads $R$. The visual message that there is no dramatic difference is corroborated by the confusion matrix in Table \ref{tab.confusion-uniformpriorbasecase}; the correct classification rate has declined only modestly, to $(1601 + 1717 + 1470)/5869 = 81.58$\%.

\begin{table}[htbp]
\begin{center}
\begin{tabular}{|l|r|r|r|r|}
\hline
 & \multicolumn{3}{c|}{Decision} &
\\
\hline
Source & Adeno & COVID & SARS & Sum
\\
\hline
Adeno & 1601  &  115 & 250 & 1966
\\
\hline
COVID &   64 & 1717 & 215 & 1996
\\
\hline
SARS  &  268 &  169 & 1470 & 1907
\\
\hline
Sum  &  1933 & 2001 & 1935 & 5869
\\
\hline
\end{tabular}
\end{center}
\caption{Confusion matrix for the MAP classifier for the case of actual reads, ``no knowledge'' priors, and correct triplet distribution-based models. The correct classification rate is 81.58\%.}
\label{tab.confusion-uniformpriorbasecase}
\end{table}

\begin{figure}[htbp]
\begin{center}
\includegraphics[width=6.5in]{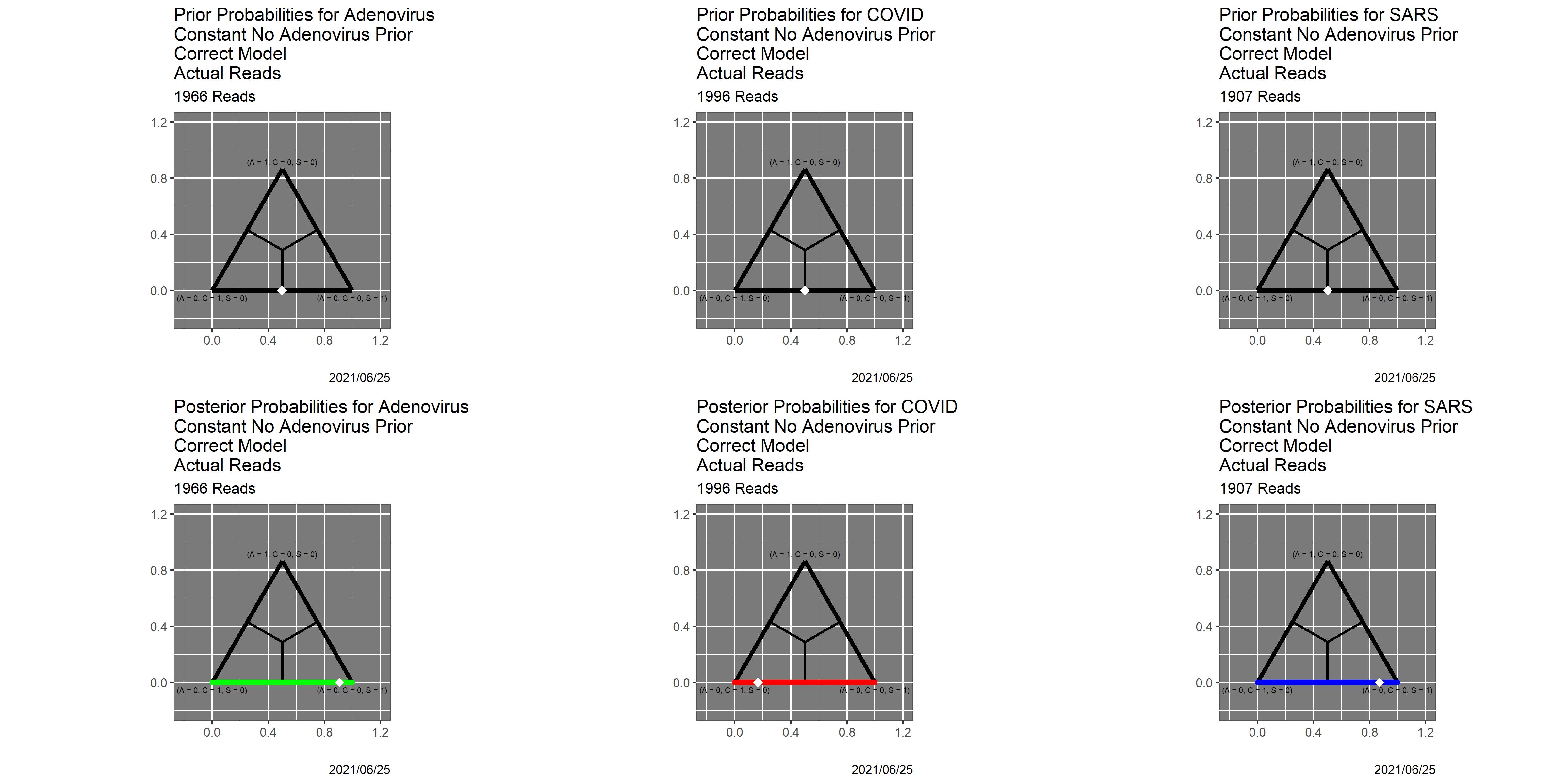}
\end{center}
\caption{Prior and posterior probabilities for the case of actual reads, ``no adenovirus priors'' and correct triplet distribution-based models. \emph{Top left:} Prior probabilities for reads from adenovirus. \emph{Top middle:} Prior probabilities for reads from COVID. \emph{Top right:} Prior probabilities for reads from SARS. \emph{Bottom left:} Posterior probabilities for reads from adenovirus. \emph{Bottom middle:} Posterior probabilities for reads from COVID. \emph{Bottom right:} Posterior probabilities for reads from SARS.}
\label{fig.bayes-noadeno}
\end{figure}

Zeros in priors, however, can be highly problematic. The ``no adenovirus'' priors $\pi_R = (0,1/2,1/2)$ instantiate a mistaken belief that no adenovirus reads are present. Figure \ref{fig.bayes-noadeno} contains results for these priors, actual reads, and the correct triplet distribution-based models. All posterior probabilities lie on the COVID--SARS axis---the bottom edge of the triangle. Table \ref{tab.confusion-noadeno} contains the associated confusion matrix. Because, as (\ref{eq.bayes}) makes clear, a zero in a prior distribution is inherited by the posterior, all adenovirus reads are misclassified, most of them as SARS. The operational lesson is that ``if it might be there, allow for it in the priors.'' See \S\ref{subsec.contamination} for further discussion in the context of contamination.

\begin{table}
\begin{center}
\begin{tabular}{|l|r|r|r|}
\hline
 & \multicolumn{2}{c|}{Decision} &
\\
\hline
& COVID & SARS & Sum
\\
\hline
Adeno  & 163 & 1803 & 1966
\\
\hline
COVID & 1728 & 268 & 1996
\\
\hline
SARS &   177 & 1730 & 1907
\\
\hline
Sum &   2068 & 3801 & 5869
\\
\hline
\end{tabular}
\end{center}
\caption{Confusion matrix for the MAP classifier for the case of actual reads, ``no adenovirus'' priors, and correct triplet distribution-based models. The correct classification rate is 58.92\%}
\label{tab.confusion-noadeno}
\end{table}

Hereafter, we consider only the informative priors, and omit prior probabilities from the graphics.

\begin{figure}[htbp]
\begin{center}
\includegraphics[width=6.5in]{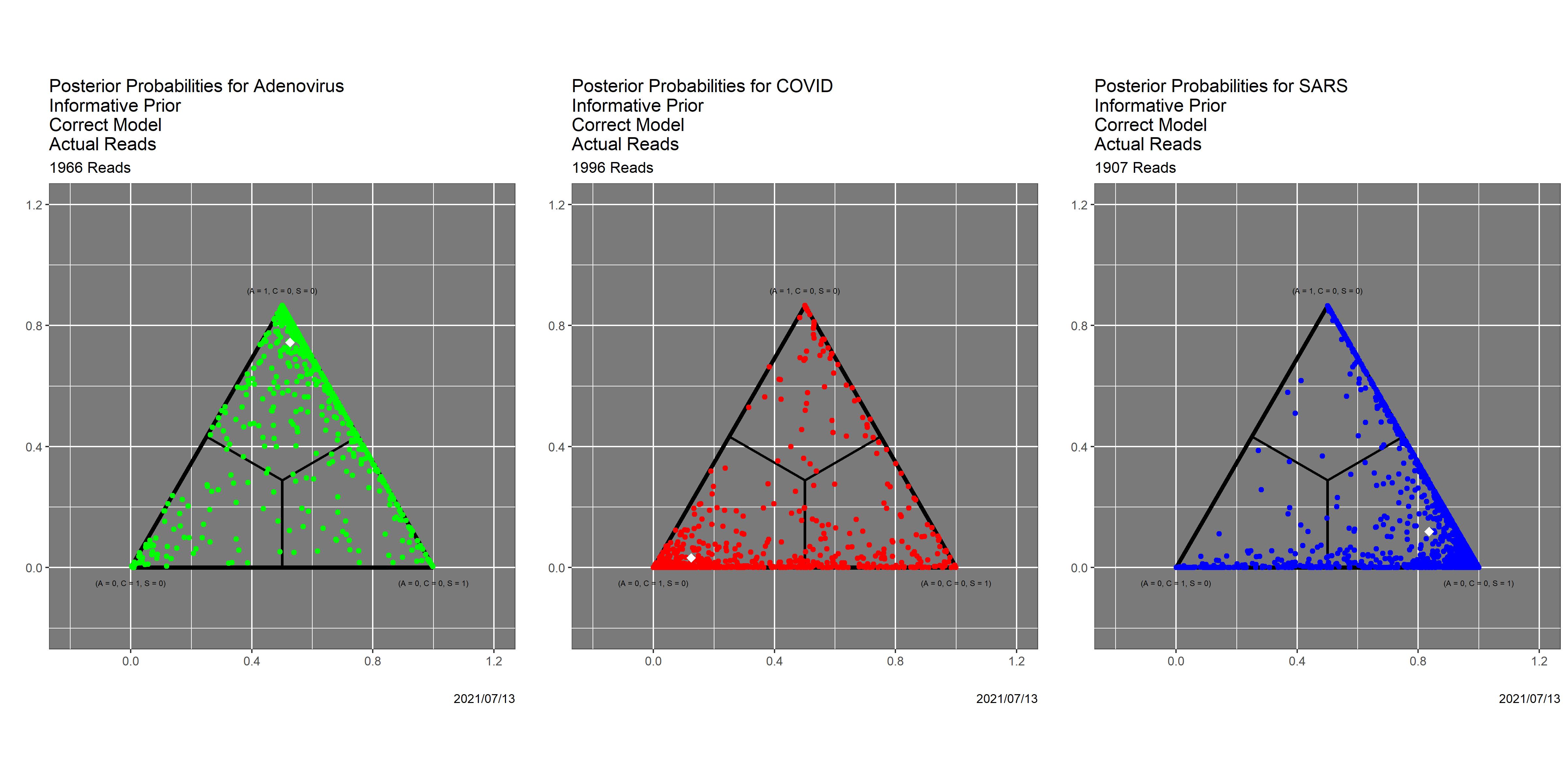}

\includegraphics[width=6.5in]{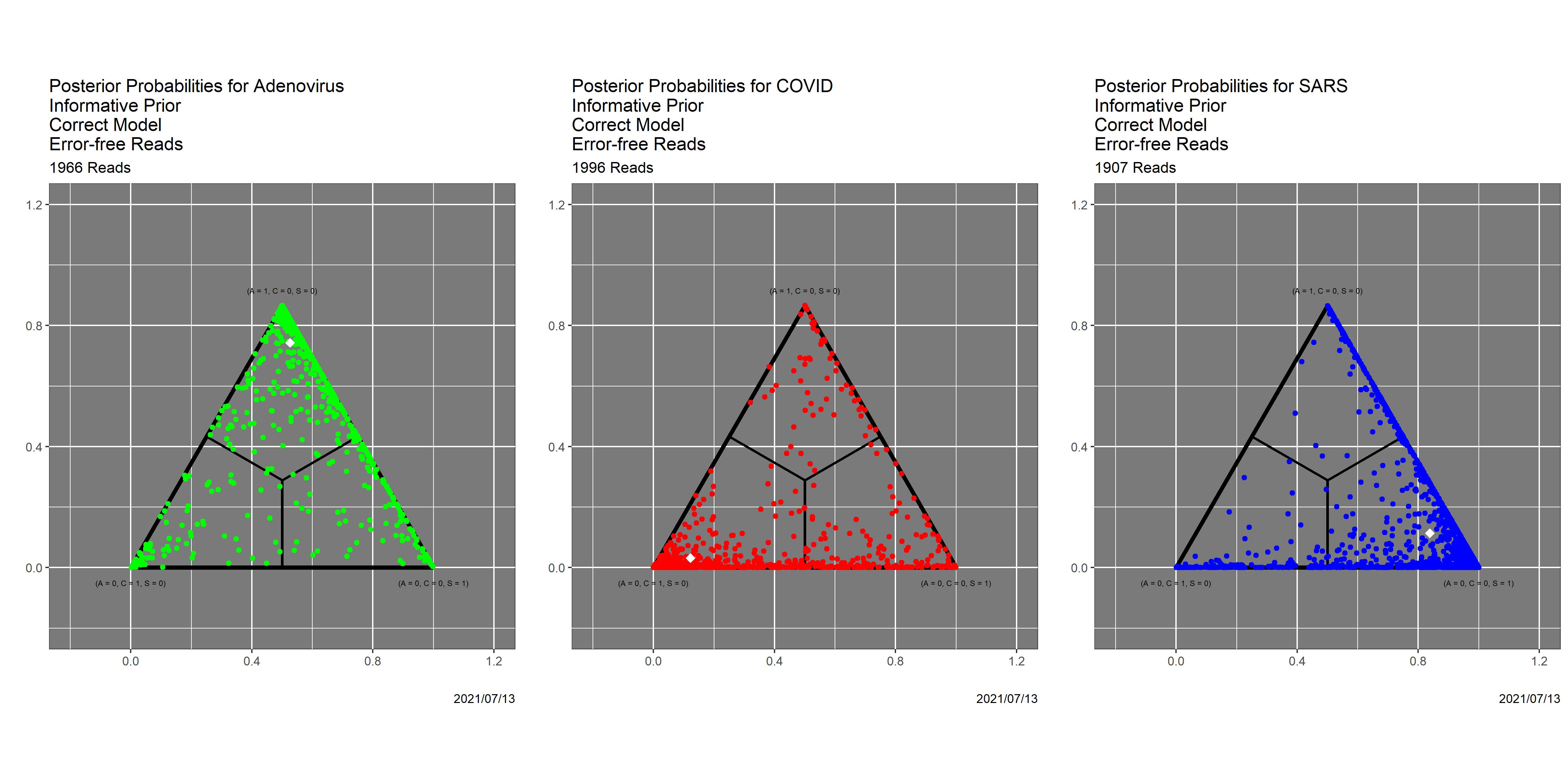}

\includegraphics[width=6.5in]{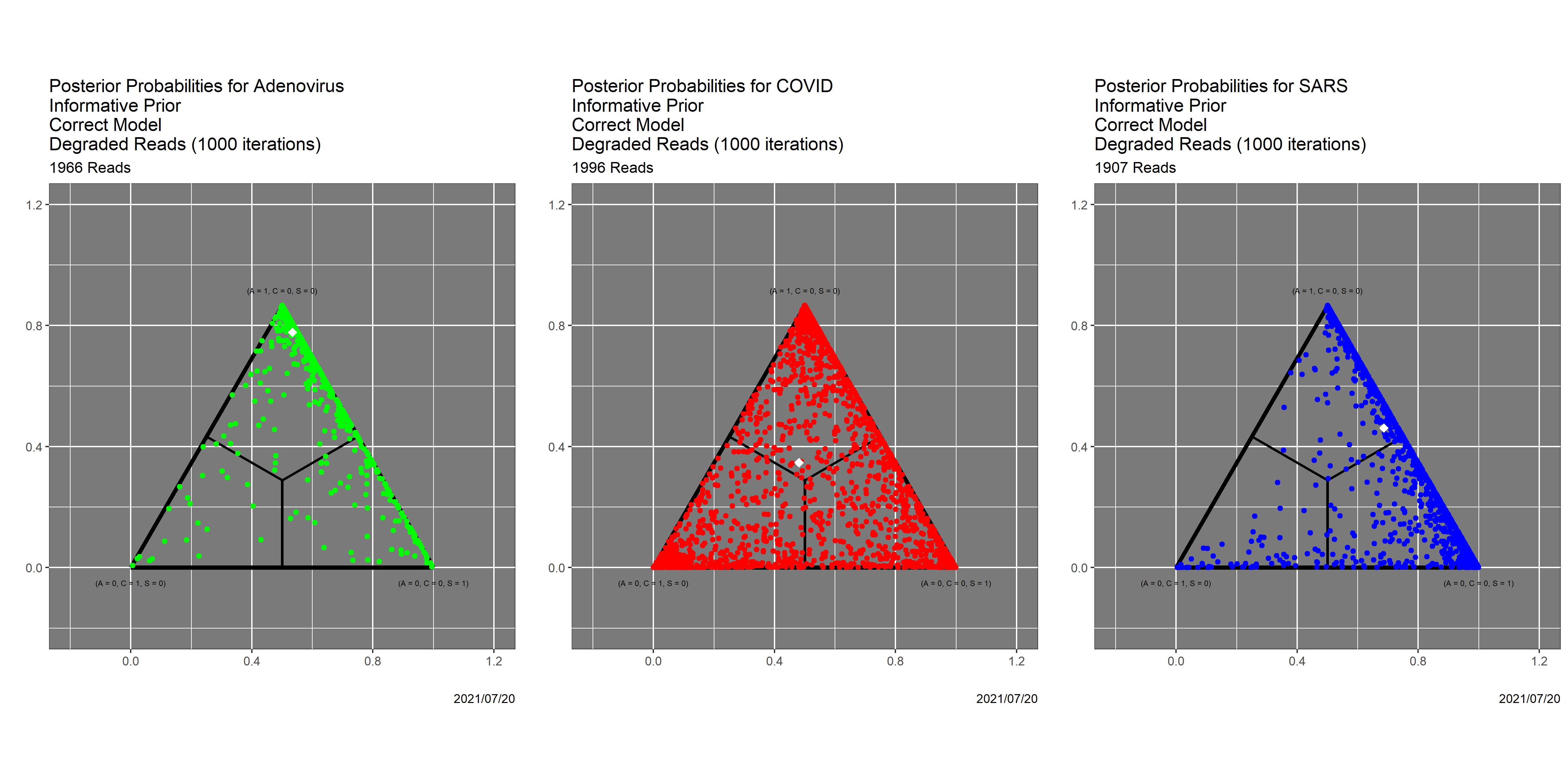}
\end{center}
\caption{Posterior probabilities when input data quality is changed. \emph{Top:} Base case (actual reads, informative priors, correct models). \emph{Middle:} Error-free reads, informative priors, correct models. \emph{Bottom:} Degraded reads, informative priors, correct models.}
\label{fig.bayes-readquality}
\end{figure}

\textbf{Changing Read (= Input Data) Quality.} Figure \ref{fig.bayes-readquality} shows the effects of increasing and decreasing the quality of the simulated reads. To enable comparisons, the top panel there is the base case posterior probabilities appearing at the bottom of Figure \ref{fig.bayes-basecase}. The middle panel shows that using the error-free reads in place of the \MS-generated reads makes no meaningful difference. This is because the default error rates in the \MS\ are not large.

More dramatic is the effect of degrading the reads using 1000 iterations of the \MV, in the manner of \cite{dqdegradation-2021}, which is shown in the bottom panel in Figure \ref{fig.bayes-readquality}. The associated MAP confusion matrix is in Table \ref{tab.confusion-degradedreads1000}. The capability to classify COVID and SARS reads correctly decreases significantly, but the same is not true for adenovirus reads: the number of correctly classified adenovirus reads is higher than in the base case.

\begin{table}[htbp]
\begin{center}
\begin{tabular}{|l|r|r|r|r|}
\hline
 & \multicolumn{3}{c|}{Decision} &
\\
\hline
       & Adeno  & COVID & SARS & Sum
\\
\hline
Adeno  & 1827   & 22    & 117  & 1966
\\
\hline
COVID  & 821    & 644   & 531  & 1996
\\
\hline
SARS   & 1042   & 65    & 800  & 1907
\\
\hline
Sum    & 3690   & 731   & 1448 & 5869
\\
\hline
\end{tabular}
\end{center}
\caption{Confusion matrix for the MAP classifier for the case of reads degraded by 1000 \MV\ iterations, informative priors, and correct triplet distribution-based model. The correct classification rate is 54.80\%}
\label{tab.confusion-degradedreads1000}
\end{table}

\begin{figure}
\begin{center}
\includegraphics[width=6.5in]{figs/PosteriorProbabilities_InformativePrior_CorrectModel_ActualReads_v2.jpg}

\includegraphics[width=6.55in]{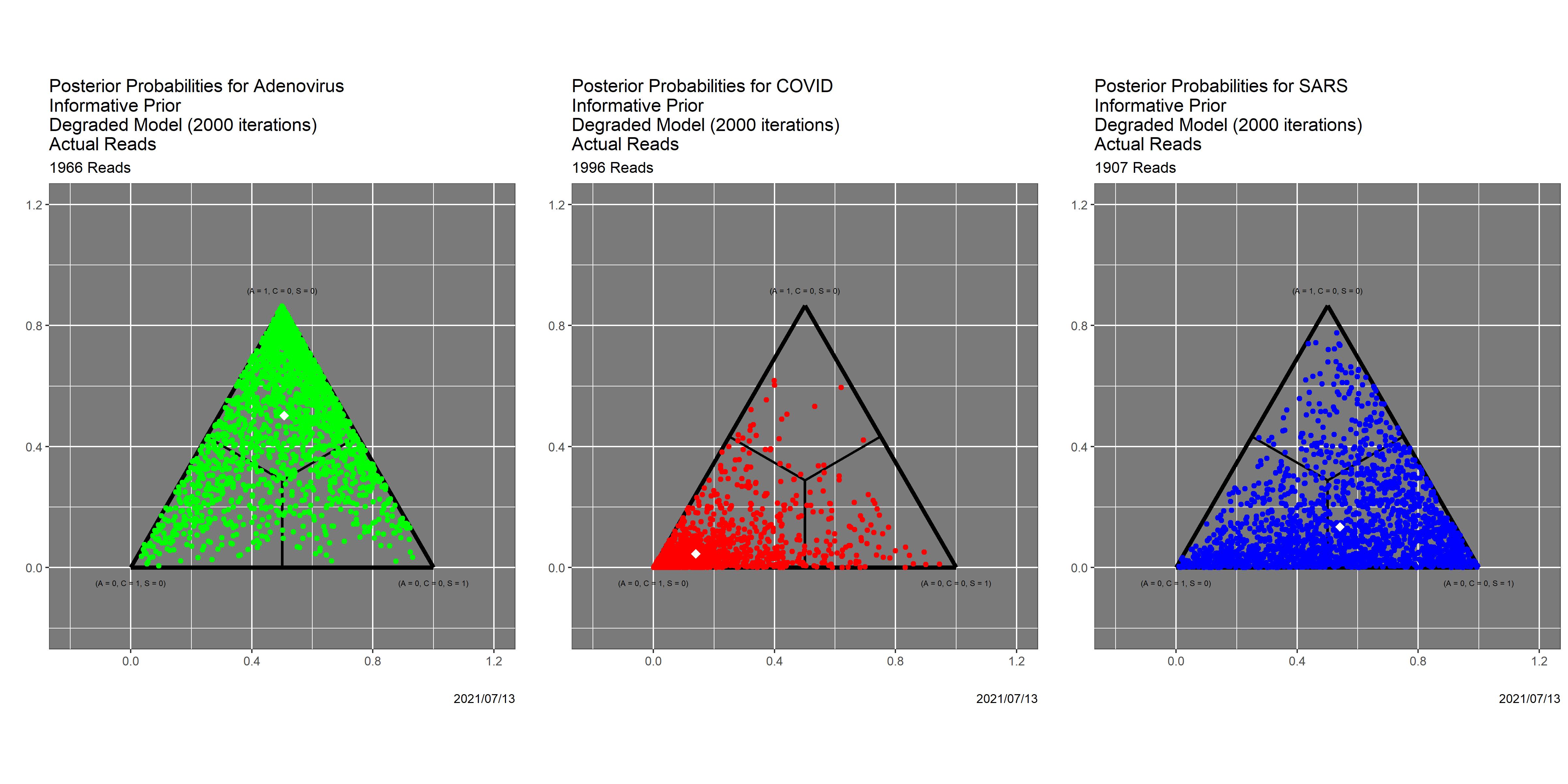}

\end{center}
\caption{Posterior probabilities showing the effect of low quality models resulting from low quality training data. \emph{Top:} Base case---actual reads, informative priors, correct models. \emph{Bottom:} Actual reads, informative priors, degraded models.}
\label{fig.bayes-degradedmodels}
\end{figure}

\textbf{Changing Model (= Training Data) Quality.} Finally, we consider the effects of decreased model quality resulting from low quality training data. Specifically, before creating the triplet distribution likelihood functions, we subject each of the three genomes to degradation by 2000 iterations of the \MV. Figure \ref{fig.bayes-degradedmodels} shows the results. As in the preceding variation, adenovirus suffers more than COVID or SARS. Comparison of the lower panel there with the bottom panel in Figure \ref{fig.bayes-readquality} shows that the effect of poor input data in the former exceeds the effect of poor training data in the latter.

\subsection{Detecting Contamination}\label{subsec.contamination}
We sketch here how our strategy can be applied to detect contamination, a major issue in some genomics and metagenomics settings.\footnote{For example, \cite{reich-2018} cites microbial and experimenter contamination as major problems in the context of ancient DNA.} To maintain continuity, we use the same three reference genomes as in \S\ref{sec.readclass}, but with adenovirus viewed as the contaminant.

\begin{figure}[htbp]
\begin{center}
\includegraphics[width=6.5in]{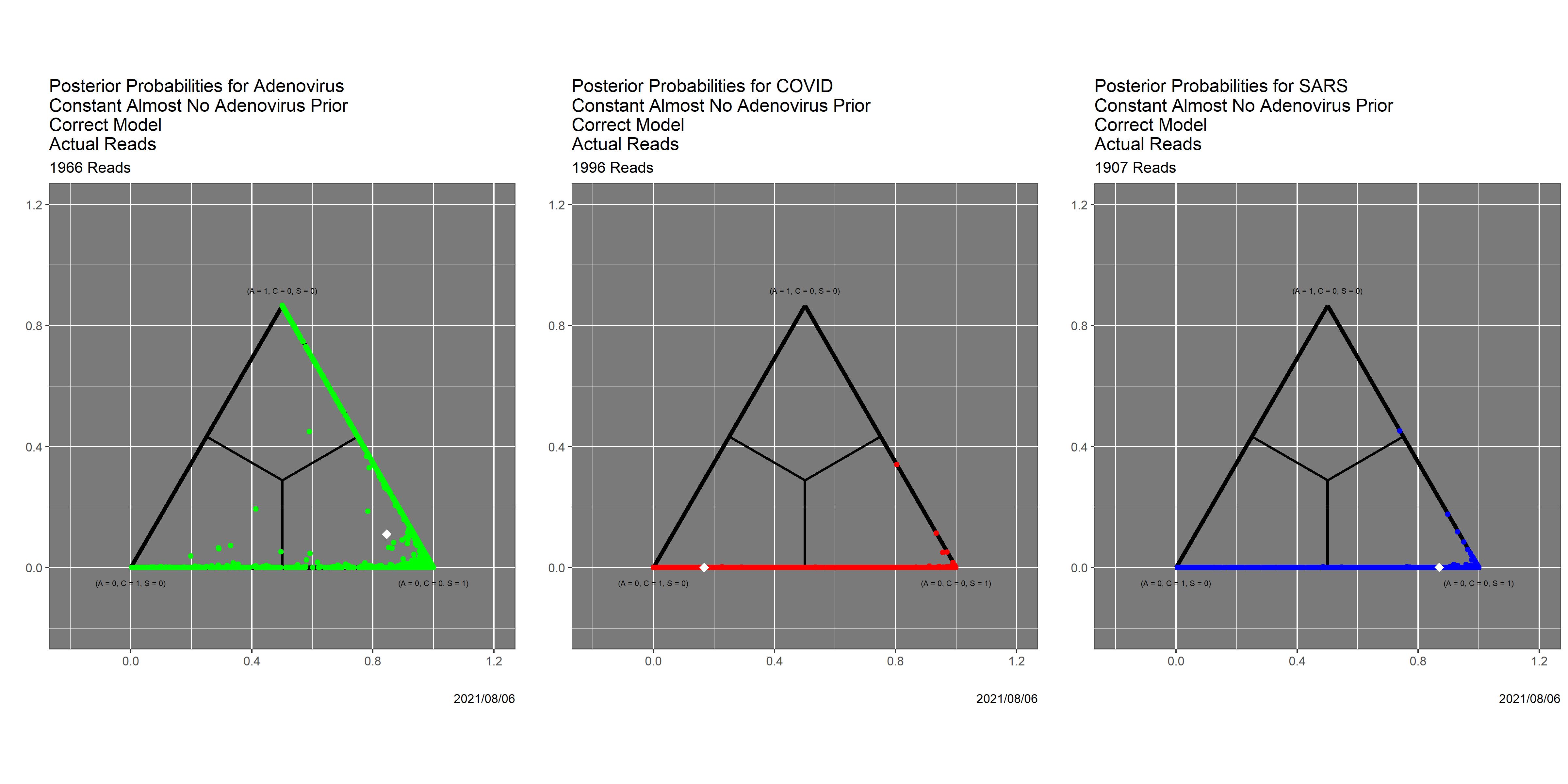}
\end{center}
\caption{Posterior probabilities for the case of actual reads, ``almost no adenovirus priors'' and correct triplet distribution-based models. \emph{Left:} Posterior probabilities for reads from adenovirus. \emph{Middle:} Posterior probabilities for reads from COVID. \emph{Right:} Posterior probabilities for reads from SARS.}
\label{fig.bayes-almostnoadeno}
\end{figure}

The key point is addressed in Figure \ref{fig.bayes-noadeno}: contamination not allowed for in the priors cannot be detected. If we replace the ``no adenovirus'' priors $\pi_R = (0, 1/2, 1/2)$ underlying Figure \ref{fig.bayes-noadeno} by ``almost no adenovirus'' priors $\pi_R = (0.0001, 0.9999/2, 0.9999/2)$, the result is as shown in Figure \ref{fig.bayes-almostnoadeno}. For the original task of classifying (roughly comparable numbers of) reads, the performance is noticeably less that in than the base case of Figure \ref{fig.bayes-basecase}, which is confirmed by comparison of the confusion matrices---that for the base case in Table \ref{tab.confusion-basecase}, and that for the case at hand in Table \ref{tab.confusion-alomstnoadeno}. The correct classification rate drops from 86.35\% to 62.57\%, which is sobering, but not frightening. The performance for COVID and SARS reads is undiminished. Crucial for what follows, the presence of adenovirus reads is confirmed.

\mbox{}
\begin{table}[htbp]
\begin{center}
\begin{tabular}{|l|r|r|r|r|}
\hline
& \multicolumn{3}{c|}{Decision} &
\\
\hline
Source & Adeno & COVID & SARS & Sum
\\
\hline
Adeno  & 215 &  163 & 1589 & 1966
\\
\hline
COVID  &   0 & 1728 & 268 & 1996
\\
\hline
SARS  &    1 &  177 & 1729 & 1907
\\
\hline
Sum  &   469 &  2063 & 3337 & 5869
\\
\hline
\end{tabular}
\caption{Confusion matrix for the MAP classifier for the case of actual reads, ``almost no adenovirus'' prior and correct triplet distribution-based models. Rows are read sources; columns are MAP decisions. The correct classification rate is 62.57\%.}
\label{tab.confusion-alomstnoadeno}
\end{center}\end{table}

Can much smaller numbers of adenovirus reads be detected by our methodology? A plausible strategy is to establish a threshold on the posterior probability of adenovirus and to declare, or at least suspect, contamination if sufficiently many---which might be as few as one---adenovirus posterior probabilities exceed that threshold. One candidate for the threshold is the maximum of the adenovirus posterior probabilities over the 3903 = 1996 + 1907 COVID and SARS reads in the dataset, which is
\begin{equation}
T^* = \max \big\{P(\textrm{Adeno|R}) : \mbox{Source}(R) \neq \textrm{Adeno} \big\} = 0.5219,
\label{eq.maxprob}
\end{equation}
where $p(\cdot|R)$ is the posterior probability given by (\ref{eq.bayes}). Note that false positives do not occur if the threshold is greater than $T^*$, so there is no benefit from using a larger threshold.

\begin{figure}[htbp]
\begin{center}
\includegraphics[width=3in]{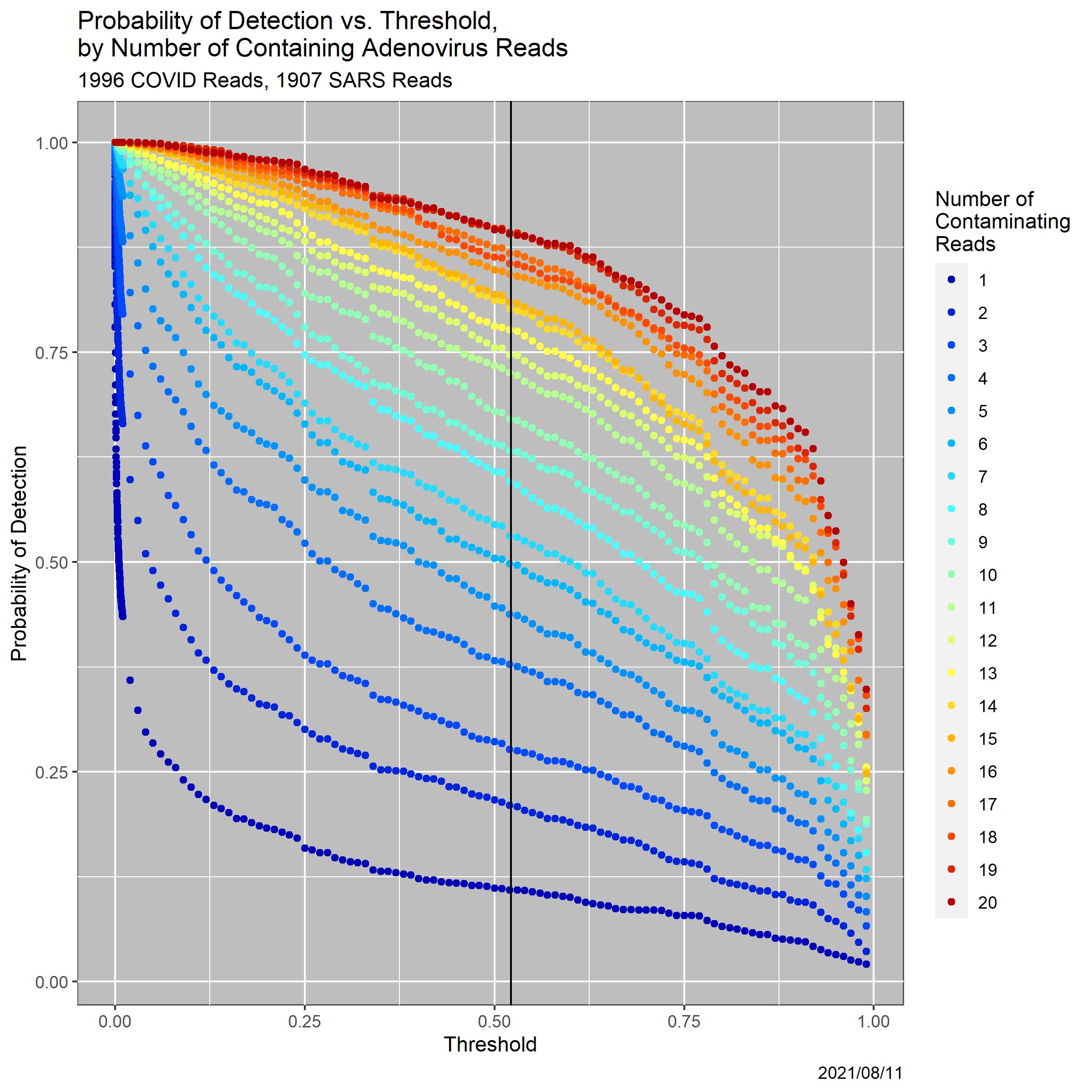}
\end{center}
\caption{Detection probabilities for adenovirus contamination as a function of threshold and number of contamination reads.}
\label{fig.detectionprobabilities}
\end{figure}

Arguably, a traditional \ROC\ curve approach \citep{fawcett-roc-2016, recordlinkage-plos1-2019} is too aggressive; while values of area under the curve (AUC) are high, unless the utility function is skewed, there are too many false positives. Figure \ref{fig.detectionprobabilities} points toward an alternative decision strategy. It shows the mean probability of detection, averaged over 2000 replicates, for numbers of contaminating adenovirus reads ranging from 1 to 20, as a function of threshold. The vertical black line is the value $T^*$ given by (\ref{eq.maxprob}). As noted previously, there are no false positives for thresholds exceeding $T^*$. For 20 reads, the probability of detecting contamination at this threshold is nearly 0.9, and even for only one read, it still exceeds 0.1.

\section{Conclusion}\label{sec.conclusion}
Triplet distributions are the ``sweet spot'' means of dimension reduction for virus genomes. In the applications to outlier identification and read classification---the major contributions of the paper, the results are not only statistically strong but also scientifically interpretable.

To understand the extent to which these results apply to longer genomes, or larger genome databases, requires further research. Computationally, there are not major impediments: almost all of the tools employed here have memory and running time demands that are linear in genome length and database size. The cluster analyses in \S\ref{sec.outliers} are an exception, because they require distance matrices whose size is the square of the database size (divided by 2, of course). The read classification problem is linear in the number of classes and the number of reads. The big question is whether, for instance, a 64- or 21-dimensional summary of a 3 million \BP\ genome (bacteria) or 3 billion \BP\ genome (humans) captures enough information to be both statistically useful and scientifically credible.

\section*{Acknowledgements}
This research was supported in part by NIH grant 5R01AI100947--06, ``Algorithms and Software for the Assembly of Metagenomic Data,'' to the University of Maryland College Park (Mihai Pop, PI).

\bibliographystyle{apalike}
\bibliography{AFK}

\appendix

\section{Example Triplet and Amino Acid Distributions}\label{app.example}
Table \ref{tab.tripletdistributions} contains the triplet distributions for the three virus genomes underlying the read classification problem in \S\ref{sec.readclass}. The corresponding amino acid distributions are in Table \ref{tab.aminoaciddistributions}. We discuss in \S\ref{sec.readclass} the extent to which these differ across the three genomes.

\begin{table}[htbp]
\begin{center}
\begin{scriptsize}
\begin{tabular}{lrrr||lrrr}
  \hline
Triplet & Adeno & COVID & SARS & Triplet & Adeno & COVID & SARS \\
  \hline
  AAA & 0.031826 & 0.026367 & 0.025581 & GAA & 0.017496 & 0.012899 & 0.015261 \\
  AAC & 0.020016 & 0.010660 & 0.018051 & GAC & 0.012220 & 0.005982 & 0.012337 \\
  AAG & 0.018463 & 0.016776 & 0.018891 & GAG & 0.013832 & 0.007252 & 0.013278 \\
  AAT & 0.018551 & 0.030176 & 0.021917 & GAT & 0.011927 & 0.021621 & 0.015597 \\
  ACA & 0.020016 & 0.012699 & 0.026219 & GCA & 0.017847 & 0.008221 & 0.014421 \\
  ACC & 0.015327 & 0.006851 & 0.013076 & GCC & 0.015063 & 0.004278 & 0.007866 \\
  ACG & 0.010316 & 0.003709 & 0.005210 & GCG & 0.015327 & 0.002373 & 0.004975 \\
  ACT & 0.016265 & 0.016609 & 0.022018 & GCT & 0.016499 & 0.013868 & 0.020908 \\
  AGA & 0.014213 & 0.016208 & 0.018085 & GGA & 0.016704 & 0.005514 & 0.011698 \\
  AGC & 0.017701 & 0.007519 & 0.011765 & GGC & 0.014243 & 0.005247 & 0.009849 \\
  AGG & 0.015591 & 0.008421 & 0.013984 & GGG & 0.012279 & 0.003409 & 0.004975 \\
  AGT & 0.014711 & 0.019015 & 0.015059 & GGT & 0.013041 & 0.018346 & 0.013916 \\
  ATA & 0.012836 & 0.024429 & 0.013345 & GTA & 0.013334 & 0.020151 & 0.014723 \\
  ATC & 0.011224 & 0.010694 & 0.011294 & GTC & 0.010228 & 0.007419 & 0.009984 \\
  ATG & 0.017378 & 0.029475 & 0.026085 & GTG & 0.014067 & 0.016308 & 0.018488 \\
  ATT & 0.018990 & 0.038765 & 0.024438 & GTT & 0.016294 & 0.037562 & 0.019698 \\
  CAA & 0.020573 & 0.012565 & 0.024471 & TAA & 0.018961 & 0.032148 & 0.019160 \\
  CAC & 0.014008 & 0.006015 & 0.016202 & TAC & 0.015679 & 0.017244 & 0.019933 \\
  CAG & 0.019342 & 0.008956 & 0.014891 & TAG & 0.010579 & 0.018179 & 0.011832 \\
  CAT & 0.017115 & 0.012030 & 0.018589 & TAT & 0.012836 & 0.039533 & 0.019026 \\
  CCA & 0.019400 & 0.006149 & 0.013311 & TCA & 0.013774 & 0.012498 & 0.020202 \\
  CCC & 0.014067 & 0.002573 & 0.004773 & TCC & 0.013744 & 0.006115 & 0.007059 \\
  CCG & 0.010257 & 0.001604 & 0.003059 & TCG & 0.008059 & 0.003676 & 0.005849 \\
  CCT & 0.014506 & 0.009491 & 0.011631 & TCT & 0.014301 & 0.019416 & 0.019093 \\
  CGA & 0.009055 & 0.002406 & 0.004672 & TGA & 0.015503 & 0.023593 & 0.022018 \\
  CGC & 0.015503 & 0.001771 & 0.004471 & TGC & 0.017290 & 0.014203 & 0.022085 \\
  CGG & 0.010257 & 0.001571 & 0.002756 & TGG & 0.018170 & 0.019115 & 0.018723 \\
  CGT & 0.009143 & 0.005614 & 0.007194 & TGT & 0.017027 & 0.038464 & 0.026724 \\
  CTA & 0.013012 & 0.017544 & 0.018723 & TTA & 0.018873 & 0.044981 & 0.023160 \\
  CTC & 0.011459 & 0.007085 & 0.012000 & TTC & 0.016968 & 0.016508 & 0.018925 \\
  CTG & 0.017525 & 0.014203 & 0.018891 & TTG & 0.019019 & 0.035390 & 0.026085 \\
  CTT & 0.019576 & 0.020552 & 0.024034 & TTT & 0.030595 & 0.059985 & 0.027463 \\
   \hline
\end{tabular}
\end{scriptsize}
\caption{Triplet distributions for the adenovirus, COVID and SARS genomes underlying \S \ref{sec.readclass}.}
\label{tab.tripletdistributions}
\end{center}
\end{table}

\begin{table}[htbp]
\begin{center}
\begin{scriptsize}
\begin{tabular}{lrrr}
  \hline
  Amino Acid & Adeno & COVID & SARS \\
  \hline
  Alanine & 0.043959 & 0.011362 & 0.019093 \\
  Arginine & 0.092782 & 0.054271 & 0.074322 \\
  Asparagine & 0.037892 & 0.080370 & 0.049245 \\
  Aspartate & 0.030537 & 0.031747 & 0.037615 \\
  Cysteine & 0.030331 & 0.016408 & 0.031430 \\
  Glutamate & 0.031035 & 0.027637 & 0.036035 \\
  Glutomine & 0.024294 & 0.023727 & 0.028472 \\
  Glycine & 0.058231 & 0.019817 & 0.032774 \\
  Histidine & 0.029628 & 0.057713 & 0.034421 \\
  Isoleucine & 0.042376 & 0.089861 & 0.050018 \\
  Leucine & 0.094042 & 0.088591 & 0.096440 \\
  Lysine & 0.047563 & 0.076494 & 0.046388 \\
  Methionine; START & 0.015679 & 0.017244 & 0.019933 \\
  Phenylalanine & 0.050289 & 0.043143 & 0.044472 \\
  Proline & 0.056267 & 0.032516 & 0.040438 \\
  Serine & 0.084049 & 0.067337 & 0.084944 \\
  STOP & 0.046479 & 0.066067 & 0.057750 \\
  Threonine & 0.067989 & 0.095375 & 0.089549 \\
  Tryptophan & 0.015327 & 0.006851 & 0.013076 \\
  Tyrosine & 0.030214 & 0.053903 & 0.039430 \\
  Valine & 0.071037 & 0.039567 & 0.074154 \\
  \hline
\end{tabular}
\end{scriptsize}
\caption{Amino acid distributions for the adenovirus, COVID and SARS genomes underlying \S \ref{sec.readclass}.}
\label{tab.aminoaciddistributions}
\end{center}\end{table}

\def\thisfile{TimesTemplate.tex}
\def\thisfiledate{2014/10/21}
\typeout{***** `\thisfile' <\thisfiledate> *****}
\end{document}